\documentclass[preprint,aps,showpacs]{revtex4}
\usepackage{amsmath}
\usepackage{graphicx}
\usepackage{epsfig}
\newlength{\upit}\upit=0.1truein
\newcommand{\raiser}[1]{\raisebox{\upit}[0cm][0cm]{#1}}
\newcommand{\ltappr}{{{\lower4pt\hbox{$<$} } \atop \widetilde{ \ \ \ }}}
\newlength{\bxwidth}\bxwidth=1.5 truein

\newcommand{\cg}{{\cal G}}

\newcommand{\str}{\hbox{Str}}
\newcommand{\Str}{\underline{\hbox{Str}}}
\newcommand{\tr}{{\hbox{Tr}}}
\newcommand{\Tr}{\underline{\hbox{Tr}}}
\begin{document}
\newcommand{\dg}{^{\dagger }}
\newcommand{\vk}{\vec k}
\newcommand{\vq}{{\vec{q}}}
\newcommand{\vp}{\bf{p}}
\newcommand{\al}{\alpha}
\newcommand{\be}{\beta}
\newcommand{\si}{\sigma}
\newcommand{\rarrow}{\rightarrow}
\def\fig#1#2{\includegraphics[height=#1]{#2}}
\def\figx#1#2{\includegraphics[width=#1]{#2}}
\newlength{\figwidth}
\figwidth=10cm
\newlength{\shift}
\shift=-0.2cm
\newcommand{\fg}[3]
{
\begin{figure}[ht]

\vspace*{-0cm}
\[
\includegraphics[width=\figwidth]{#1}
\]
\vskip -0.2cm
\caption{\label{#2}
\small#3
}
\end{figure}}
\newcommand{\fgb}[3]
{
\begin{figure}[b]
\vskip 0.0cm
\begin{equation}
\includegraphics[width=\figwidth]{#1}
\end{equation}
\vskip -0.2cm
\caption{\label{#2}
\small#3
}
\end{figure}}
%\psdraft
\newcommand\frmup[1]{\raiser{\epsfig{file=#1,width=\bxwidth}}}

\newcommand \bea {\begin{eqnarray} }
\newcommand \eea {\end{eqnarray}}
\newcommand{\bk}{{\bf{k}}}
\newcommand{\bx}{{\bf{x}}}

\title{Sum Rules and Ward Identities in the Kondo Lattice}

\author{P. Coleman$^{1,2}$, I. Paul$^{1,3}$ and J. Rech$^{1,2,3}$ }
\affiliation{$^{1}$Kavli Institute for Theoretical Physics, Kohn Hall, UCSB
Santa Barbara, CA 93106, USA} 
\affiliation{
$^2$Center for Materials Theory,
Rutgers University, Piscataway, NJ 08855, U.S.A. } 
\affiliation{$^{3}$ SPhT, L'Orme des Merisiers, CEA-Saclay, 91191
Gif-sur-Yvette France.}
%\date{}
\begin{abstract}
We derive a generalized
Luttinger-Ward expression for the Free energy 
of a many body system involving a constrained Hilbert space.
In the large $N$ limit, we are able to explicitly write the entropy as
a functional of the Green's functions. 
Using this method we obtain 
a Luttinger sum rule for the Kondo lattice.
One of the fascinating aspects of the sum rule, is that it contains 
two components, one describing the heavy electron Fermi
surface, the other, a sea of oppositely charged, spinless fermions.
In the heavy electron state, 
this sea of spinless fermions is completely filled and the electron Fermi
surface expands by one electron per unit cell to compensate the
positively charged background, forming 
a ``large'' Fermi surface.
Arbitrarily weak magnetism causes the 
spinless Fermi sea to annihilate with part of the Fermi sea of the
conduction electrons, leading to a small Fermi surface.
Our results thus enable us to show that the Fermi surface volume 
contracts from a large, to a small volume at a quantum critical point. 
However, the sum rules also permit the possible 
formation of a new phase, sandwiched between the
antiferromagnet and the heavy electron phase, where the charged spinless
fermions develop a true Fermi surface. 
\end{abstract}

%\eject
%
\maketitle
%
%\vfill\eject 

\section{Introduction}\label{Intro} 

Sum rules  play a vital  role in
condensed matter  physics.  The most  famous sum rule -  the Luttinger
sum  rule\cite{luttinger,luttinger2,luttinger3}, defines rigorously  the volume  of the  Fermi surface  of a
Fermi liquid  in terms of  the density of electrons: 
\begin{equation}\label{luttsumrule}
2 \frac{{\rm v}_{FS}}{(2\pi)^{D}} = n_{e}
\end{equation}
where v$_{FS}$ is the Fermi surface volume,
$n_{e}$ the density of electrons per unit cell and $D$ the dimension.
Historically, sum
rules  have also  played an  important  role in  our understanding  of
strongly correlated systems. In the context of the Kondo effect
for example, the  Friedel sum rule\cite{friedel,langer,langreth} 
\begin{equation}\label{friedsumrule}
\sum_{\lambda}\frac{\delta_{\lambda}}{\pi} = \Delta n_{e}
\end{equation}
relating the sum of the scattering phase shifts in channels labelled
by $\lambda$ to the number of bound-states $\Delta n_{e}$ helped to 
establish a rigorous foundation for the Abrikosov-Suhl resonance\cite{abrikosov,suhl,gruner} which develops
in Anderson and Kondo impurity models. 
Later,
Martin\cite{martin} applied the  Luttinger sum rule
to the Anderson  lattice model, to  argue that  heavy electron
metals must have a ``large Fermi surface'' which counts both the
conduction electrons and also the localized f-electrons.
\begin{equation}\label{largeFS}
2\frac{{\rm v}_{FS}}{(2\pi)^{D}} = n_{e}+1.
\end{equation}

Today, there is a renewed interest
in sum rules, in connection with models of strongly correlated
electrons.
For instance, in the context of high temperature superconductors,
which are Mott insulators when undoped, 
there has been a long-standing debate over whether the
``large'' Fermi surface predicted by Luttinger's sum rule, might be
replaced at low doping,  by a ``small'' Fermi surface determined by the number of
doped holes\cite{schraiman,chubukov}. Related issues arise in the context of heavy electron systems, where
experimental advances
have made it possible to tune through the 
quantum critical point that separates the heavy electron paramagnet
from the local moment antiferromagnet.
There have been a number of theoretical speculations that the Fermi surface
of the heavy electron material may jump from ``large'' to ``small''
at the quantum critical point\cite{coleman,senthil,pepin05}. Recent experimental work, based on de
Haas van Alphen measurements\cite{onuki} and Hall
measurements\cite{silke}, provide experimental support
for this hypothesis, but the idea has lacked rigorous theoretical support.

The main difficulty in extending sum rules to strongly correlated
systems, is that the theoretical machinery 
used and developed by Luttinger and Ward to derive sum rules 
applies to models with un-projected
Hilbert spaces. About eight years ago Affleck and
Oshikawa\cite{affleck} demonstrated that such sum rules have a more
general
existence. By  using
a modification of the Leib Mattis theorem, Affleck and Oshikawa showed
that the ``large Fermi surface'' which counts both local
moments and conduction electrons develops in the one dimensional $S=1/2$
Kondo model, even though in this case, the ground-state is not a Fermi liquid.
More recently, Oshikawa\cite{oshikawa} has extended this
derivation to higher dimensional Kondo lattices. This work suggests
that it ought to be possible to extend the Luttinger Ward approach to
Hamiltonian problems with strong constraints.

In this paper, we show
how the original methods of Luttinger and Ward can indeed be extended to
strongly correlated models. 
A key element of this work is the construction 
of a functional relating the Free energy of a strongly
interacting system to the Green's functions of the projected Hilbert space. Our approach
is based on the use of 
slave particles, such as a Schwinger boson description for the local
moments in a Kondo lattice. 
We begin with a 
generalization of the Luttinger Ward Free energy functional,
appropriate for systems of interacting bosons and
fermions\cite{eliashberg,blaizot} which can be compactly written as 
\begin{equation}\label{notsobigdeal}
F = T{\Str}\left[\ln  \left(- {\cal G}^{-1} \right)+
 \Sigma {\cal
G} \right]+ Y[{\cal G}]
\end{equation}
where 
 $\Str[A] = \Tr[A_{B}]-\Tr[A_{F}]$ denotes the supertrace of a
matrix containing both bosonic (B) and fermionic (F) components
(where the underline notation is used to denote a sum over internal
frequencies and a trace over the internal quantum numbers 
of the matrix). $\cg 
= (\cg _{0}^{-1}- \Sigma )^{-1}$ 
is the matrix describing the fully dressed Green's 
function of all elementary particles and fields entering the
Lagrangian, including the slave particles, where  $\Sigma $ is the
self-energy 
matrix and ${\cal G }_{0}$ the bare propagator of the fields. The quantity
$Y[\cg]$ is, diagramatically, the sum of all closed-loop two-particle
irreducible skeleton
Feynman diagrams (Fig. \ref{fig1}.).  \figwidth=14cm
\fg{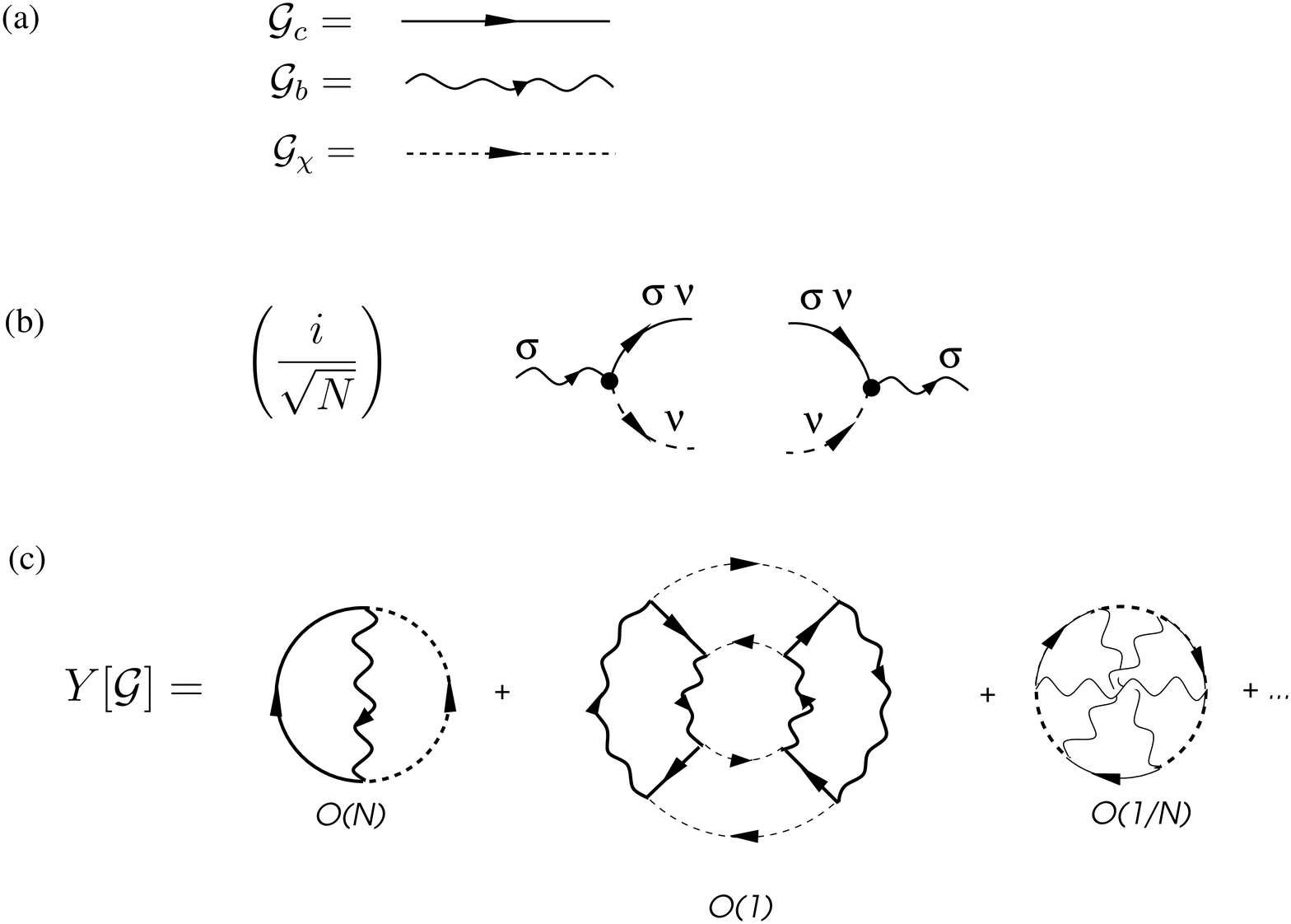}{fig1}{Illustrating the Luttinger Ward functional for the Kondo
lattice. (a) Fully renormalized  propagators for the conduction electrons, 
$\chi $ fermions and Schwinger bosons, where $\sigma\in [1,N] $ is the spin
index and $\nu\in[1,K] $ the channel index,  (b) Interaction vertices
showing spin ($\sigma $) and channel ($\nu$) indices and
(c)
leading skeleton diagrams for the Luttinger Ward functional with 
dependence on $\frac{1}{N}$. The first diagram involves two loops
carrying spin and channel quantum numbers, and one pair of vertices and is
consequently of order $O (KN/N)= O (N)$. The second diagram has four
loops carrying internal quantum numbers, and four pairs of vertices,
so it is of order $O (N^{2}K^{2}/N^{4})= O (1)$. The final diagram has two
quantum number loops and three pairs of vertices, and is hence of
order $O (NK/N^{3})\sim O (1/N)$.
}
Variations of $Y[\cg ]$ with
respect to the Green's function $\cg $ generate the self energy
\begin{equation}\label{eq:varyx}
\delta Y[\cg] = -T\Str \left[\Sigma \delta \cg \right].
\end{equation}
or
\begin{equation}
\Sigma (\omega)=- \beta \frac{\delta Y}{\delta \cg (\omega)} 
\end{equation}
where the use of the supertrace as the measure for functional derivatives
avoids the need to introduce a relative minus sign between Fermi and
Bose parts of this expression. 
Diagrammatically, the functional derivative of $Y$ with respect to $\cg
$ corresponds to ``cutting'' one of its internal lines. If we truncate
$Y$ to some order in $1/N$, this relationship determines a conserving
Kadanoff-Baym approximation\cite{kadanoffbaym}. To leading order in the large $N$
expansion, the self-energies generated from the functional derivatives
of $Y$ are
\bxwidth=2in
\upit=-0.35truein
\begin{eqnarray}\label{largeNself}
\cr
\Sigma_{b} (\omega)&=&-\beta \frac{\delta Y}{\delta \cg_{b} (\omega)}
=\ \  \frmup{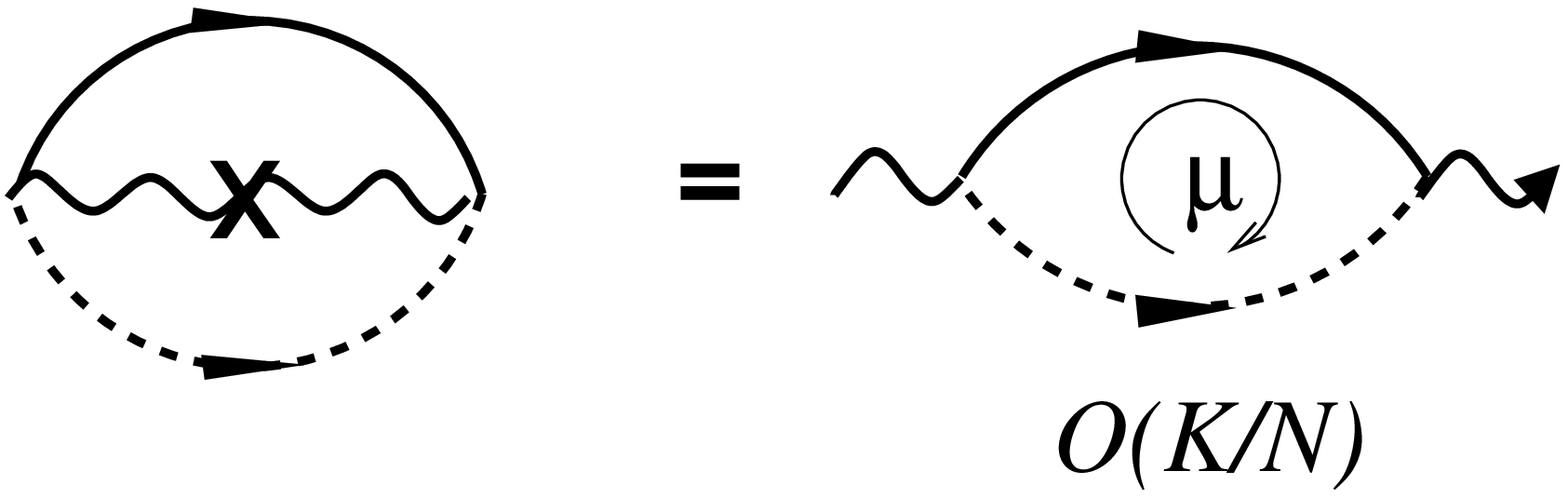}\ \ ,
\cr\cr\cr
\Sigma_{\chi } (\omega)&=&+\beta \frac{\delta Y}{\delta \cg_{\chi }
(\omega)} 
= \ \  \frmup{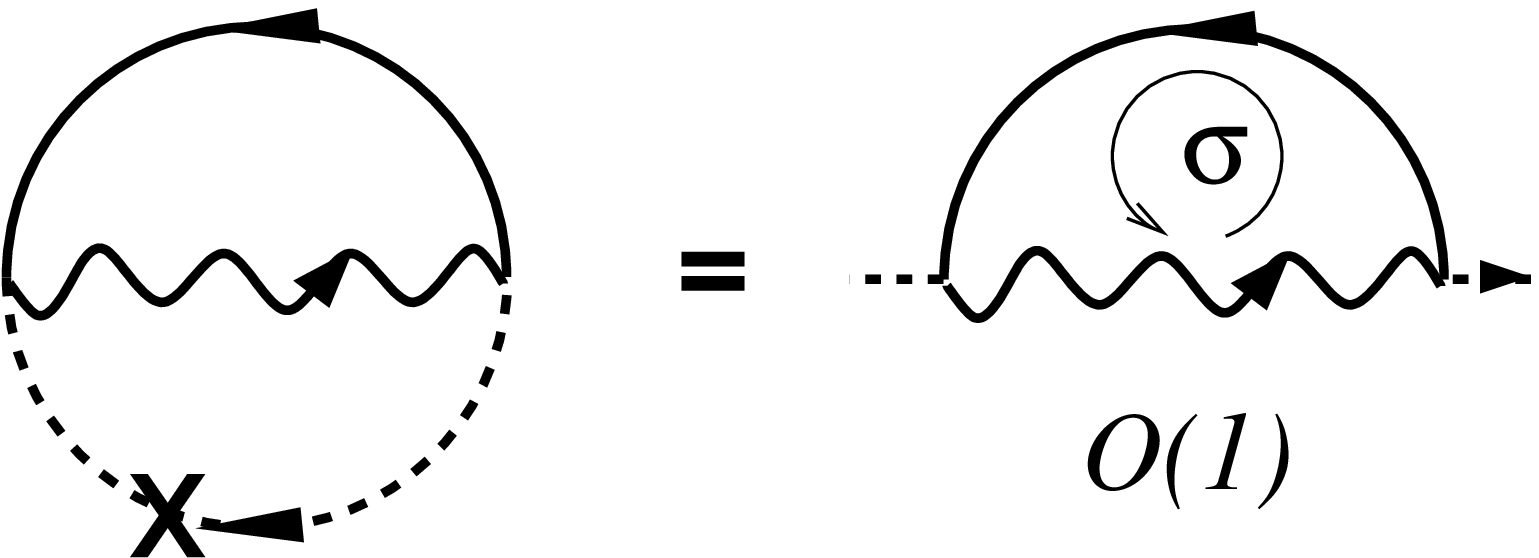}\ \ ,
\cr\cr\cr
\Sigma_{c} (\omega)&=&+\beta \frac{\delta Y}{\delta \cg_{c} (\omega)}
= \ \  \frmup{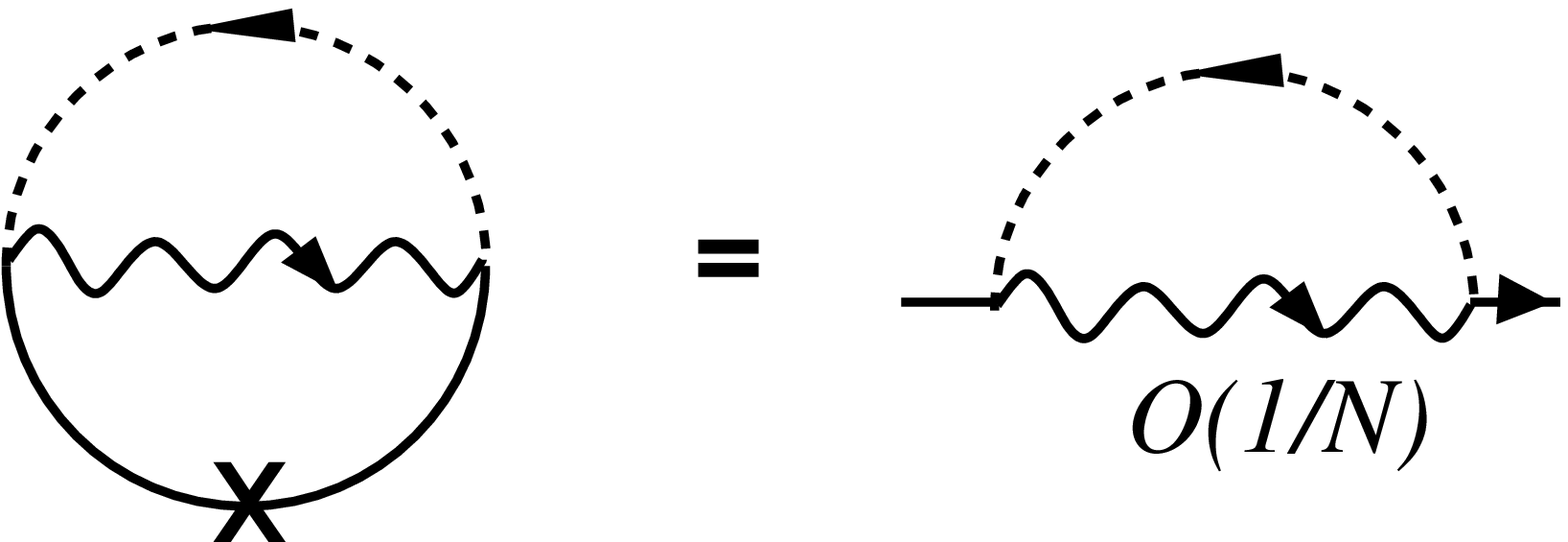}\ \ ,
\end{eqnarray}
\vskip 0.3in
\noindent where the cross indicates the line which is eliminated by
the functional differential. 
Each of these terms contains a factor $O (1/N)$ from the vertices, but
the first two self-energies contain summations over the  internal
channel or spin indices, elevating $\Sigma_{b}$ and $\Sigma_{\chi }$
to terms of order $O (1)$. In the leading order large $N$
approximation, $\Sigma_{c}\sim O (1/N)$, so that a consistent large N
approximation is produced by leaving the conduction electron lines un-dressed.
This provides an alternative diagrammatic derivation of the Parcollet-Georges
approach to the multi-channel Kondo model\cite{parcollet97a}.

Each conserved quantity $Q$ which commutes
with the Hamiltonian leads to a gauge invariance of the underlying
quantum fields, such that the action and all physical properties
are invariant under the transformation
\begin{equation}\label{gaugetransform}
\psi_{\zeta}
 (t) \rightarrow e^{i \theta (t) q_{\zeta}}\psi_{\zeta}.
\end{equation}
Here, $q_{\zeta }$ is the gauge charge of the field:
it is this quantity which controls the charge of any
physical excitations associated with the field. 
Provided this gauge invariance is unbroken  we show that this leads to
a Ward identity at zero temperature
\begin{equation}\label{WardId}
\int_{-i\infty}^{i\infty}\frac{d\omega}{2 \pi i} \str \left[\hat
q \cg \frac{d\Sigma (\omega)}{d\omega} \right]=0.
\end{equation}
where the integral runs along the imaginary axis. Luttinger's original
work derived the elementary version of this 
relationship for electrons: here it is seen to hold for interacting
species of fields.

Each of these Ward identities leads to a corresponding, 
generalized Luttinger sum rule
given by
\begin{equation}\label{generalLutt}
Q=\frac{1}{\pi} Im 
{\rm Tr}\left[ \phantom{\int}\hskip -0.1in 
\hat  q 
\ln \left( 
- {\cg}_{F}^{-1} (0-i\delta )\right) \right]
\end{equation}
where $\cg_{F}$ is the complete Fermionic Green's function, involving
all fermions involved in the description of the constrained Hilbert
space, including those introduced as slave particles. In impurity
systems, the trace over the logarithms reduces to a sum of phase shifts,  and
the quantity $Q$ is replaced by $\Delta Q$, the change in $Q$ induced
by the impurity. In the limit of infinite band-width, $\Delta Q=0$,
which leads to a multi-particle version of 
Friedel sum rule
\begin{equation}
0 = \sum q_{\zeta}n_{\zeta} \left( \frac{\delta_{\zeta}}{\pi}\right)
\end{equation}
where $n_{\zeta}$ is the spin degeneracy associated with the trace
over internal quantum numbers of the field $\psi_{\zeta}$. 
In lattice
systems, the trace over the log becomes the Fermi surface
volume associated with the field $\psi_{\zeta}$, so that 
\begin{equation}
Q= \sum_{\zeta} \left[\phantom{\int}
q_{\zeta}n_{\zeta}
\frac{{\rm  v}_{FS} ({\zeta})
}{(2  \pi)^{D}} \right],
\end{equation}
where 
${\rm  v}_{FS} ({\zeta})$ is the Fermi
surface volume associated with field $\psi_{\zeta}$. A closely related
Fermi surface sum rule has recently been derived by Powell et al. 
for mixtures of fermions and bosons in atom traps\cite{powell}. 

In this paper we apply these results to the Kondo lattice model. 
Key to our approach is the adoption of Schwinger bosons 
for the description of quantum spins. 
Schwinger bosons offer a key advantage, because they can describe
both the antiferromagnetic and the 
heavy electron ground-states of the Kondo lattice\cite{rech}. 
The Kondo effect induces a 
development of a retarded interaction in the spin singlet channel
between the electrons and the spins.  Formally, this 
interaction manifests itself as the mediating field $\chi_{j}$
in the  Hubbard-Stratonovich decomposition
of the interaction\cite{parcollet97a,colemanpepin03,indranil}
\begin{equation}
J_{K}\vec{ S}_{j}\cdot c\dg_{j\alpha }
\vec{\sigma}_{\alpha
\beta}c_{j\beta }
\rightarrow  
\frac{1}{\sqrt{N}}\left(c\dg_{j\alpha }b_{j\alpha}   \right)
\chi\dg _{j}
+ 
\frac{1}{\sqrt{N}}\left(b\dg _{j\alpha }c_{j\alpha } \right)
\chi _j + \frac{\chi\dg _{j}\chi_{j}}{J_{K}}
\end{equation}
Here $\vec{S}_{j}$ is the spin
at site $j$ and $c\dg _{j\alpha} $ creates an electron at site j. 
At 
long-times the $\chi $  field 
develops charge dynamics which describe
the scaling of the Kondo interaction in both frequency and momentum
space. In particular, the $\chi -$ propagator 
describes the momentum and frequency dependent Kondo interaction,
\begin{equation}
-{\cg}_{\chi} (\bk ,\omega) = \left( \frac{1}{J_{K}}+ \Sigma_{\chi}
(\bk ,\omega)\right)^{-1} = J^{*}_{K} (\bk, \omega ).
\end{equation}
At each point in momentum space where 
the Kondo interaction scales to strong coupling, ${\cg}_{\chi }$ develops
a pole. In this way, the physics of the
Kondo effect is intimately linked\cite{pepin05} to the possible emergence of a
positively charged, spinless Fermi field - a ``holon''\cite{zouanderson}. 

Under gauge transformations, the conduction and $\chi $ fields
have opposite gauge charge,
\begin{equation}
q_{c}=1, \qquad q_{\chi}=-1.
\end{equation}
When we apply the Luttinger Ward procedure,
we find that the physical  charge density is given by the total volume of
the electron Fermi surface, minus the volume of the holon Fermi
surface, as follows
\begin{equation}
n_{e}= 2 \frac{{\rm v}_{FS}}{(2 \pi)^{D}} -  \frac{{\rm v}_{\chi }}{(2 \pi)^{D}}
\end{equation}
Here, the volume of the electron Fermi surface is given by the
conventional Luttinger formula,
\begin{equation}
{\rm v}_{FS}= \frac{1}{\pi} \sum_{\bk }{\rm Im}\ln [\epsilon_{\bk}+
\Sigma_{c} (\bk ,0-i \delta)-\mu + i \delta  )] = \sum_{\bk}\Theta (\mu-E_\bk)
\end{equation}
where $\mu$ is the chemical potential, 
$E_{\bk }= \epsilon_{\bk }
+ \Sigma_{c} (\bk, E_{\bk })$ is the renormalized energy of the heavy
electrons. Now  ${\rm v}_{\chi }$ is 
\begin{equation}
{\rm v}_{\chi } = \frac{1}{\pi} \sum_{\bk }{\rm Im}\ln [\frac{1}{J_{K}}+
\Sigma_{\chi } (\bk ,0-i \delta))]= \sum_{\bk }\Theta (-J^{*}_{K} (\bk ))
\end{equation}
where  $J^{*}_{K} (\bk )= -{\cg}_{\chi} (\bk ,0-i \delta)$.

The meaning of the renormalized coupling constant $J^{*}_{K}$ in the
paramagnetic phase needs a little discussion. This quantity
describes the residual interaction between the electron
fluid and any additional spins that are added to the Fermi liquid ground-state. 
If we add an
additional Schwinger boson, increasing $S\rightarrow S+\frac{1}{2}$ at a given site, the
additional spin unit remains unscreened, because each channel can only
screen one spin unit, and all channels are fully screening.  If
$\omega_{b}= E (n_{b}+1)-E (n_{b})$ is the energy to adding one
additional spin, then the residual interaction between the additional
Schwinger boson and the conduction electrons is $J_{K}^{*}
(\omega_{b})$. Since the additional spin decouples, it follows that
$J_{K}^{*} (\omega)$ has a zero at $\omega=\omega_{b}$, 
$J_{K}^{*} (\omega_{b})=0$. At
higher energies, the residual interaction will become
positive, ultimately connecting up to the single-ion scaling behavior
$J_{K}^{* } (\omega)\sim 1/ \ln  (T_K/\omega)$. Since $J_{K}^{*}$
passes through zero at $\omega=\omega_{b}$, it follows that at lower energies $\omega<\omega_{b}$, and in
particular, at zero frequency, the residual coupling must be
ferromagnetic, i.e $J_{K}^{*} (0) <0$. 
This in turn, means that the Fermi sea of holons is entirely full 
${\rm  v}_{\chi }= (2 \pi)^{D}$, so that 
\begin{equation}
n_{e}= 2 \frac{{\rm v}_{FS}}{(2 \pi)^{D}} - 1.
\end{equation}
To preserve the overall charge density, the electron 
Fermi surface volume is forced to enlarge by one unit per spin
to ``screen'' the finite background density of postively charged
holons (Fig. \ref{fig5}.).

Let us now consider the antiferromagnet. In this case, 
the Schwinger boson field
condenses, and the $\chi  $ field becomes hybridized with the
conduction electrons.  In this case, 
the holon Fermi surface annihilates with the expanded
part of the heavy electron Fermi surface, to 
reveal a single integrated set of Fermi surface sheets
with a ``small'' volume that encloses the total electron count.  If the transition
from paramagnet to antiferromagnet occurs via a single quantum
critical point, then our results indicate that the transition
in the Fermi surface volume at the second order quantum critical
point is  abrupt. \cite{ambiguity}
\fg{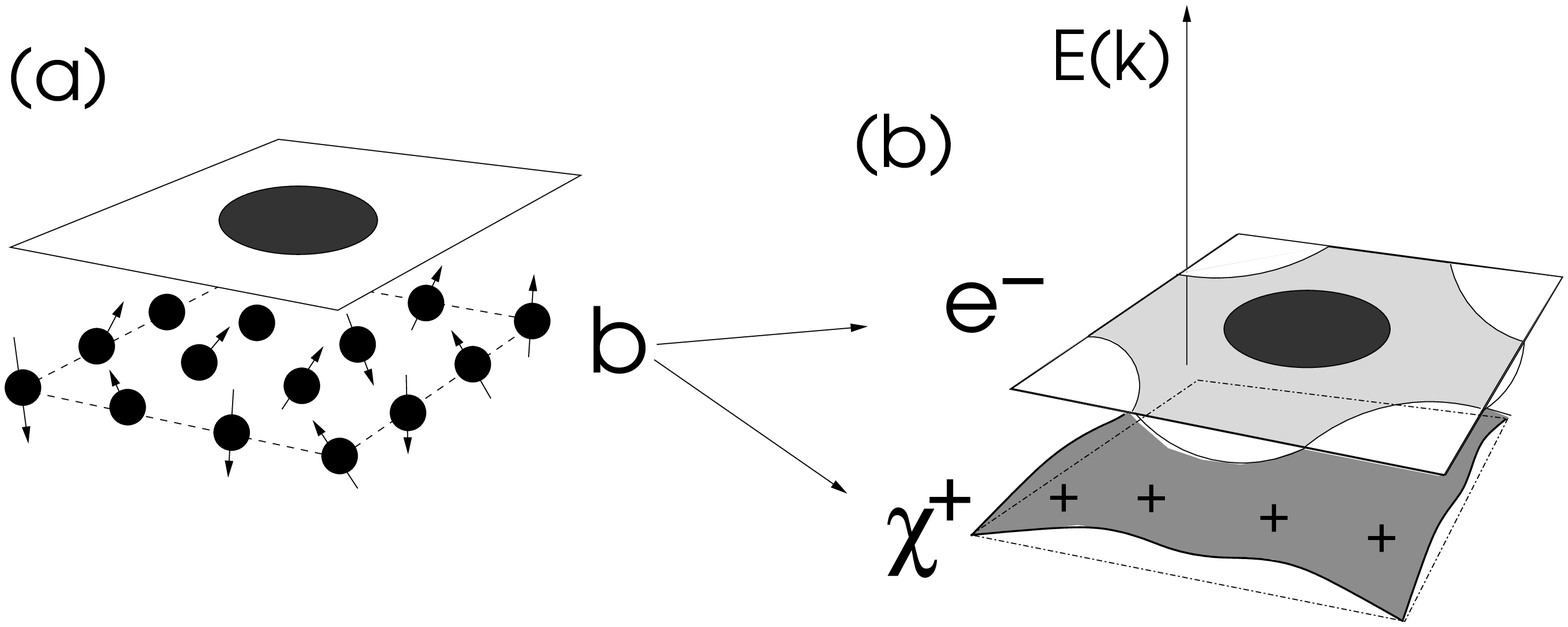}{fig5}{Schematic diagram illustrating the
enlargement of the Fermi surface in the Kondo lattice (a) coupling of
local moments to small Fermi surface (b) each spin-boson ``ionizes''
into an electron and a ``holon'', producing a large electron Fermi
surface and a filled holon Fermi sea. }

Perhaps the most interesting aspect of this picture, is that it suggests
that the background field of holons can develop dynamics.
One intriguing possibility, is that the holons are liberated 
when the Fermi surface volume contracts at the quantum critical
point.
Another more exotic possibility, allowed by the Ward identities,
is the formation of a 
separate {\sl phase}
where the holons develop a Fermi surface, so that ${\rm v}_{\chi }
= x (2 \pi)^{D}$ is partially filled.  
This  ``spin-charge'' decoupled phase would be
sandwiched between the localized magnetic phase and the fully formed
paramagnet. P\' epin has recently suggested that the holons might 
develop a Fermi surface at the quantum critical point\cite{pepin05}.
This might provide a mechanism for such a scenario. 
A good candidate for this phase is the one-dimensional
Haldane $S=1$ chain,
coupled to two conducting chains to form a one-dimensional, two-channel Kondo lattice. 
Here, the conjectured intermediate phase would seperate a Haldane gap phase and a large
Fermi surface phase, leading to 
Friedel oscillations with wavevector $q =  2 k_{F}^{(0)}+
\frac{\pi}{2}x $,
intermediate 
between that of the large ($x=1$) and small ($x=0$) Fermi surface.

This paper is divided up into various parts. In the first we develop the
Luttinger-Ward Free energy expansion for the 
Kondo lattice. Next, we present a new formula for the entropy of the
interacting fluid, expressed exclusively in terms of the Green's
functions of the particles. We then go on to derive the Luttinger sum rule in a general form for a constrained
system, which we then apply to the Kondo
impurity and lattice model. In the final part, we discuss
the general implications of our sum rule in the Kondo lattice.

\section{\bf{Luttinger Ward Expression for the Kondo Lattice}}

The Luttinger Ward approach was originally based on perturbation theory
in the strength of the Coulomb interaction.  In order to develop
these methods for strongly interacting systems, we need to take
account of the constraints in the theory, requiring a Luttinger Ward approach
appropriate to gauge theories.  In this respect, our efforts to apply the
Luttinger Ward approach to the Kondo lattice closely parallel recent efforts
to extend the Luttinger Ward approach to quark-gluon plasmas\cite[]{blaizot}.

Although the Luttinger
Ward functional may be derived non-perturbatively\cite{potthoff}, 
the usefulness of any derived sum rules 
rests on the direct relationship between the Green's functions that
enter in the functional, and physical excitations. 
Strongly
interacting systems do not generally provide a natural small parameter on which to base
a perturbation expansion. Our philosophy is that 
to develop 
this  control, we need to define a family
of large-N models which, in the infinite $N$ limit provide a 
mean-field description of the phases of interest. In the expansion
about this limit, $1/N $ then provides
a small paramater. The large $N$ solutions constructed this way, 
will of course satisfy the sum rules we derive, but the 
sum rules are expected to have a far greater generality. 

In the context of our interest in the Kondo lattice, we adopt
a family of models 
given by
\begin{equation}
\mathcal{H} = \sum_{\vk, \nu, \al} \epsilon_{\vk} 
c^{\dg}_{\vk \nu\al} 
c_{\vk \nu \al} 
+ 
\overbrace {
\frac{J_{K}}{N}
\sum_{j \nu \al \beta} c^{\dg}_{j \nu\al} 
c_{j \nu \be}  b\dg_{j\beta}b_{j\alpha }
}^{H_{I}}
\end{equation}
Here, we have adopted the Schwinger boson
representation, where $b\dg _{j\alpha}
$ creates a Schwinger boson at site $j$, with spin component
$\alpha\in [1,N]
$. The  combination
$b^{\dg}_{j \al} b_{j \be}={S} _{\al \be} (j) 
$ represents the local spin operator and 
the system is
restricted to the physical Hilbert space by requiring that 
\begin{equation}
\sum_{\al} b^{\dg}_{j\al} b_{j \al} = 2 S,
\end{equation}
at each site $j$.  The index $\nu\in [1,K]$ is a channel index that is absent in the
physical $SU (2)$ model, but which is included to create a family of
large $N$ models. The ``filled shell'' case $2S=K$ 
defines a perfectly screened
Kondo lattice, and our discussion will focus on this case. 
Here $c^{\dg}_{\vk \al} $ creates a conduction electron 
of  momentum state $\vk$ and spin $\alpha \in
(1,N)$. $c\dg _{j\alpha } = \frac{1}{\sqrt{{\cal  N}_{s}}}\sum c\dg _{\bk \alpha} e^{-
i \vec{ k} \cdot {\vec x}_{j}}$ creates an 
electron at site $j$, where ${\cal N}_{s}$ is the number of sites in
the lattice.  
The above model is known to have a controlled large $N$ expansion
which contains both magnetic and paramagnetic solutions when $N$ is
taken to infinity with $k=K/N$ fixed.

To develop a Luttinger Ward expansion, we shall factorize the interaction
in terms of a Grassman field $\chi_{j}$, 
\begin{equation}
H_{I}
\rightarrow 
\sum_{j \nu\al}
\frac{1}{\sqrt{N}}\left[( c\dg_{j\nu\alpha }b_{j\alpha} )
\chi\dg _{j\nu}
+ 
(b\dg _{j\alpha }c_{j\nu\alpha } )
\chi _{j\nu}\right] + \sum_{j\nu}\frac{\chi\dg _{j\nu}\chi_{j\nu}}{J_{K}}.
\end{equation}
and the Lagrangian becomes 
\begin{eqnarray}
{\cal L} &=& 
\sum_{\vk \al \nu} c^{\dg}_{\vk \nu \al} 
\left( \partial_{\tau} + \epsilon_{\vk} \right) c_{\vk \nu\al}
+ \sum_{j \al} b^{\dg}_{j\al} \left( \partial_{\tau} + \lambda_{j} \right)
b_{j \al} + \sum_{j\nu}\frac{\chi\dg _{j\nu}\chi_{j\nu}}{J_{K}}.\cr
&+& \sum_{j \nu\al}
\frac{1}{\sqrt{N}}\left[( c\dg_{j\nu\alpha }b_{j\alpha} )
\chi\dg _{j\nu}
+ 
(b\dg _{j\alpha }c_{j\nu\alpha } )
\chi _{j\nu}\right]- 
\sum_{j} 2S \lambda_{j}
\end{eqnarray}
Note that the Grassman fields $\chi_{j}$ contain no time derivative,
and so instantaneously, they behave as inert, neutral fields. 
Here $\lambda_{j}$ is the static value of the lagrange multiplier
used to enforce the constraint on each lattice site.
At finite temperatures the constraint is only enforced on the average,
but in the ground-state, the value of the conserved quantities 
$n_{bj}$ will be quantized as a step-function
of the $\lambda_{j}$. In the range $\lambda_{j}\in [\lambda_{j}^{-},
\lambda_{j}^{+}]$, ( where $\lambda_{j}^{\pm}$ are the energy gaps between the
physical ground-state and the states with $2S\pm 1$ Schwinger bosons
at site j)
the constraint will be precisely satisfied. We 
shall assume that a translationally invariant choice $\lambda_{j}=\lambda$
is  always available. 

For our sum rules, we need to know comparatively little about the
nature of the ground-state excitations. However, from the 
the large $N$ solution of this model, we do know that there 
are two main classes of solution to the Kondo lattice
\begin{itemize}

\item Heavy electron phase, where the Schwinger bosons pair-condense in either the Cooper channel 
($\langle \sigma b\dg _{i\sigma}b\dg _{j-\sigma}\rangle\neq 0 $
antiferromagnetic interactions)  or particle-hole channel ($\langle
\sigma b\dg _{i\sigma}b _{j\sigma}\rangle\neq 0 $
ferromagnetic interactions). In field-theory language, this is a
``Higg's phase'' in which the local $U (1)$ symmetry is broken
and both the spin and holon excitations can propagate from site to
site. 

\item Magnetic phase, in which the Schwinger bosons individually condense ($\langle
b_{j}\rangle  \ne 0 $) to develop long-range magnetic order. In this
phase, the $\chi_{j}$ fields become hybridized with the conduction fields,
and $\cg_{c}$ and $\cg_{\chi }$ become diagonal members of a single 
matrix propagator.

\end{itemize}
In general, we will therefore have to preserve the momentum dependence
of the propagators. The bare propagators of the theory are diagonal in
momentum, and given by
\begin{eqnarray}\label{a}
{\cg}_{b}^{(0)}
 (\vk,i\nu_n) &=& (i\nu_n\underline{\tau_{3}} - \lambda\underline{1})^{-1},\cr
{\cg}_{c}^{(0)}
 (\vk,i\omega_n) &=& (i\omega_n - \epsilon_{\vk })^{-1},\cr
{\cg}_{\chi }^{(0)}
 (\vk,i\omega_n) &=& - J_{K}
\end{eqnarray}
where we have written the bare boson propagator as a
two-dimensional Nambu propagator, to allow for the possible
introduction of boson pairing, which develops as antiferromagnetic
correlations begin to grow. 
Formally, we will combine these three propagators into a single
propagator given by
\begin{equation}
\cg ^{-1}= \left[
\begin{matrix}
i\nu_n\underline{\tau_{3}} - \lambda\underline{1}
&&\cr & i\omega_{n}
-\epsilon_{\vk }&\cr&&-\frac{1}{J_{K}}
\end{matrix} \right]- \Sigma 
\end{equation}
where $\Sigma $ is a matrix self-energy. Following  Luttinger and Ward, we
will regard $\cg $ as a variational function, and $\Sigma $ as a derived
quantity.

We now consider the effect of tuning up  the strength of the Kondo interaction
from zero to $J_{K}$, by replacing 
$J_K \rightarrow  \alpha J_K$ where $\alpha \in [0,1]$, keeping the chemical potential of
the conduction electrons and bosons fixed. Now the partition function
is given by
\begin{equation}
Z = Tr[e^{- \beta H}] = 
\int {\cal D}
[ c,b,\chi ] e^{- \int_{0}^{\beta}{\cal L}d\tau}
\end{equation}
where ${\cal  D}[c,b,\chi] $ is the measure of the path
integral. If we vary $\alpha$ inside this expression, we obtain
\begin{eqnarray}\label{b}
\frac{dZ}{d\alpha} & =& \int {\cal D}
[ c,b,\chi ] \sum_{j \nu}\int_{0}^{\beta}d\tau  \left(
\frac{1}{\alpha^{2}J_{K}}\chi\dg _{j \nu}\chi_{j \nu}\right)
e^{- \int_{0}^{\beta}{\cal L}d\tau }\cr
&=&
\beta Z \times 
\frac{1}{ \alpha^2 J_K} \sum_{j \nu}
 \langle\chi \dg _{j \nu}\chi_{j \nu}\rangle
\end{eqnarray}
so that if $F= - T\ln Z$ is the free energy, then
\begin{equation}
\label{firstderivative}
\frac{d F}{d \alpha} = -\frac{1}{( \alpha^2 J_K)}
 \sum_{j \nu}
 \langle\chi \dg _{j \nu} \chi_{j \nu}\rangle
= -\frac{ T}{( \alpha^2 J_K)}
{{\Tr}}
\left[ {\cg}_{\chi} \right],
\end{equation}
where ${\cg}_{\chi } $ is the Green's function for the $\chi $ fermion
and ${\Tr} [{\cg}_{\chi }] \equiv  K\sum_{i\omega_{n}, \vk }{\cg}_{\chi } (\vk ,i\omega_{n}) e^{i \omega_n 0^+} $
denotes a trace over the  frequency and momentum 
of ${\cg}_{\chi }$ (here, $e^{i \omega_{n}0^{+}}$
ensures the normal ordering of the operators).
We next consider the expression
\begin{equation}
\label{eq:interim-gnd-pot}
\tilde{F} = T \Str  \left[
\ln \left( - \cg ^{-1} \right) 
+ \Sigma \cg  \right] + Y[\cg ].
\end{equation}
We will show how this expression generalizes the Luttinger
Ward Free energy functional to a mixture of interacting bosons and
fermions. 
Here, we use the notation
\begin{equation}
\Str[{A}] \equiv  {\Tr}[A_{B}]-{\Tr}[A_{F}] \equiv  \sum_{i \nu_n}\tr[A_{B} (i
\nu_n)]e^{i \nu_n 0^{+}}
-\sum_{i \omega_n}\tr[A_{F} (i \omega_n)]e^{i \omega_n 0^{+}}
\end{equation}
to denote the ``supertrace'' 
over the bosonic and fermionic parts of the matrix $A$. The
``underline'' notation is used to indicate a sum over both the 
frequency variable  of $A$ and a trace over the internal quantum
numbers of $A$ (such as momentum).
The supertrace includes
a relative minus sign for the
fermionic component of the matrix, and we have explicitly displayed
the trace over the  discrete Matsubara
bose ($i\nu_n=(2n) \pi T$) and fermi ($i \omega_{n}= (2n+1)\pi T$)
frequencies. In the above expression, the self energy matrix $\Sigma =
{\cg}_{0}^{-1}-\cg ^{-1}$. 
The quantity 
$Y[\cg]$ is the  sum of all closed,  two-particle irreducible skeleton
Feynman graphs for the Free energy. 

As mentioned earlier, the Luttinger Ward functional $Y[\cg]$ has the property that its
variation with respect to $\cg$ generates the self energy matrix, 
\begin{equation}\label{eq:vary}
\delta Y[\cg] = -T\Str \left[\Sigma \delta \cg \right].
\end{equation}

The variation of the first term in $\tilde{F}$ with respect to $\cg$ is 
given by
\begin{equation}
\delta \left(T \Str 
\left[ \ln \left( - \cg^{-1} \right) 
\right]
 \right) = T \Str \left[ - \cg^{-1}\delta \cg\right]
\end{equation}
By using the relation $\Sigma [\cg] = {\cg}_0^{-1} - \cg^{-1}$, the variation
in the second term in $\tilde{F}$ is given by
\begin{equation}
\delta \left(T \Str 
\left[  \left(  [{\cg}_{0}^{-1}- \cg^{-1}]\cg
 \right) \right]
 \right) = T \Str \left[  {\cg}_{0}^{-1}\delta \cg\right]
\end{equation}
so that 
the total variation of $\tilde{F}$ with respect to $\cg$ 
\begin{equation}
\delta \tilde{F} = T \Str \left[\left(
\overbrace {-\cg^{-1}+{\cg}_{0}^{-1}}^{\Sigma }
- \Sigma  \right)\delta \cg \right]= 0
\end{equation}
identically vanishes, 
\begin{equation}
\frac{\delta \tilde{F}}{\delta \cg} = 0.
\end{equation}
Now the Hubbard Stratonovich transformation that we have carried out on the interaction $H_{I}$ assures that the only place that the coupling constant
$\alpha J_{K}$ enters, is in $[{\cg}_{\chi}^{0}]^{-1}= -\frac{1}{\alpha J_{K}}$.
This means that in the r.h.s of Eq.~\eqref{eq:interim-gnd-pot} 
$\alpha$ enters explicitly only through 
$\Sigma_{\chi} = -(\alpha J_K)^{-1} - [\cg_{\chi}]^{-1}$. Then,
\begin{eqnarray}
\frac{d \tilde{F}}{d \alpha}
&=& \underbrace{\frac{\delta \tilde{F}}{\delta \cg}}_{= 0} 
\frac{\partial \cg}{\partial \alpha} + \frac{\partial \tilde{F}}{\partial
\alpha}
\nonumber \\
&=&
-\frac{ T}{( \alpha^2 J_K)} {\rm \Tr} 
\left[ {\cg}_{\chi} (\vq, i \omega_n) e^{i \omega_n 0^+} \right].
\end{eqnarray}
But by comparison with (\ref{firstderivative}), we see that $\frac{d
\tilde{F}}{d \alpha}=\frac{d F}{d \alpha}$ and 
since, $F(\alpha =0) = \tilde{F}(\alpha =0)$
in the non-interacting case, the two quantities must be equal for all
$\alpha$, i.e
\begin{equation} 
\label{eq:gnd-pot}
F = 
T \Str \left[
\ln \left( - \cg^{-1} \right) + \Sigma \cg \right] 
+ Y[\cg].
\end{equation}

There are various points to make about this derivation:
\begin{itemize}

\item The derivation is very general. Its 
correctness only depends on the stationarity of
$\tilde{F}[\cg]$ with respect to variations in $\cg$ and the equivalence between
$dF/d\alpha$ and $d\tilde{F}/d\alpha$.
This means that the above
expression will hold for broken symmetry or ``Higg's phase'' 
solutions that involve off-diagonal components
to $\cg$\cite{eliashberg}. 
In the context of the Kondo lattice, this means that $\cg$ can be
extended to include  anomalous boson pairing terms that are driven by
short-range antiferromagnetic correlations, 
or alternatively,
off-diagonal terms driven by long-range magnetism, 
which mix the conduction and $\chi $ fermions.
This expression can also be used to describe superconducting
states, where the conduction electron and $\chi $ propagators will contain
off-diagonal terms.

\item Even though the free energy is a functional of three Green's functions,
there is no overcounting. 

\item The above Free energy functional can be used as a basis for
developing conserving approximations that generalize the Kadanoff Baym
approach to a constrained system\cite{kadanoffbaym}. In particular, in the large $N$
limit, the skeleton graph expansion for $Y$ truncates at the leading diagram
(Fig. \ref{fig1}.), providing the basis for a controlled treatment of both the
magnetic and the paramagnetic phases of the Kondo lattice\cite{parcollet97a,rech}.

\end{itemize}

\section{Entropy Formula}

In this section, we derive an approximate formula for the entropy of a
Luttinger Ward system which becomes exact in the large $N$ limit, and
in any approximation where the vertex corrections can be neglected. 
The result, which we shall derive 
below, is
\begin{equation}\label{entropy}
S (T, \cg ) = \int \frac{d\omega}{\pi} \Tr_{B,F} \left[\left(\frac{d \hat n
(\omega)}{dT} \right)
\left({\rm Im} \ln \left[- \cg^{-1} (\omega-i\delta ) \right]+ {\rm
Im}\Sigma (\omega) {\rm Re}\cg  (\omega)
 \right)
 \right]
\end{equation}
where
\begin{equation}\label{c}
\hat n (\omega) = \left[\begin{matrix}n (\omega)& \cr & f (\omega)\end{matrix} \right]
\end{equation}
is the matrix containing the Bose or Fermi-Dirac distribution
functions 
$n (\omega)= [e^{\beta \omega}-1]^{-1}$, $f (\omega)= [e^{\beta
\omega}+1]^{-1}$. 
The trace is carried out over the Bose and Fermi components
of both expressions. Here $\cg (\omega)$ and $\Sigma (\omega)$ are the self-energies
that have been analytically continued onto the real axis.  This
expression is extremely useful, for it only involves the low -energy
part of the Green's functions of the particles, and does not involve the functional $Y[\cg
]$. 
A version of this formula was first quoted in the context of lattice
gauge theory by Blaizot et al.\cite{blaizot}. 

Key to our approach, is the notion that the Luttinger Ward functional
$F[\cg ]$ can be rewritten in terms of the {\sl real} frequency
Green's functions. This has the advantage that one does not have to
deal with the temperature dependence of the Matsubara frequencies. 
The approach that we adopt is in fact, reminiscent of the Keldysh approach
for non-equilibrium systems\cite{rammer,kamenev}. 
To preserve the
symmetry between bosons and fermions, it proves useful to introduce
a kind of ``Keldysh'' notation, writing
\begin{equation}
\hat h (\omega)= \frac{1}{2}+ \hat \eta \hat n (\omega) =
\left[\begin{matrix}\frac{1}{2}+n (\omega)& \cr & \frac{1}{2}-f (\omega)\end{matrix} \right]
\end{equation}
Use of this function preserves the ``supertrace'' symmetry of our approach.
Consider the first part of the Free energy
\begin{equation}
F_{1}=T \str \left[
\ln \left( - \cg^{-1} \right) + \Sigma \cg \right] .
\end{equation}
We replace the summation in this term by an integral over $\hat  n
(\omega)$ around the imaginary axis,
\begin{equation}
T\Str[A]\rightarrow -\int \frac{d\omega}{\pi } {\rm Im}
\str[\hat  h (\omega) A(\omega-i\delta)]
\end{equation}
where $\str [A] = {\rm Tr}[A_{B}]- {\rm  Tr}[A_{F}]$ is the supertrace
over the spatial and internal quantum numbers of $A$.
By distorting the contour around
the real axis, we obtain
\begin{equation}
F_{1}= -
\int \frac{d\omega}{ \pi  }{\rm Im}\ \str \left[\hat  h (\omega) \left\{
\ln \left( - \cg^{-1} \right) + 
(\cg_{0}^{-1}  -\cg^{-1})
\cg \right\}_{\omega-i\delta } 
\right].
\end{equation}
If we vary the Green's function, then we obtain
\begin{equation}
\delta F_{1} = -
\int \frac{d\omega}{ \pi  }{\rm Im}\str\left[ \hat  h
(\omega)\tilde\Sigma (\omega)
\delta \cg 
\right]_{\omega-i\delta } .
\end{equation}
where we have employed the notation $\tilde{\Sigma }= \cg_{0}^{-1}- \cg^{-1}$
In a similar fashion, when we vary $\cg $ inside $Y[\cg]$, we obtain
\begin{equation}\label{eq:deltay}
\delta Y= 
\int \frac{d\omega}{ \pi  }{\rm Im}\ \str\left[ \hat  h (\omega)
\ {\Sigma} (\omega)\ \delta \cg (\omega)
\right]_{\omega-i\delta } .
\end{equation}
The condition that the
two terms cancel sets $\cg_{0}^{-1}-\cg^{-1} = \Sigma $, defining
both the real and the imaginary parts of
the self-energies in terms of the Green's functions.

We can exploit the stationarity $\delta F/\delta \cg =0$
to simplify the differentiation of 
the Free energy with respect to the temperature. When we differentiate
F, we can neglect the temperature dependence of the spectral functions.
\begin{equation}
S (T)= - \frac{dF}{dT} = 
 - \left. \frac{\partial F}{\partial
T}\right|_{\cg }\end{equation}
The contribution to the entropy from $F_{1}$
is then 
\begin{eqnarray}\label{d}
S_{1}= 
\int \frac{d\omega}{ \pi } {\rm Im}\  \str
\left(\frac{\partial\hat  h
(\omega)}{\partial T}
\left[
\ln \left( - \cg^{-1} (\omega-i\delta ) \right) + \Sigma
(\omega-i\delta ) \cg (\omega-i\delta )\right]  \right).
\end{eqnarray}
The temperature derivative of the second term $Y[\cg ]$in
(\ref{eq:gnd-pot}) requires more careful consideration. 
The general diagrammatic contribution to $Y$ contains a certain number of
frequency summations. When we analytically continue these frequency
summations, distorting the contour integrals around the branch-cuts of
the Green's functions, 
we pick up ``on-shell'' contributions from each branch cut of the form
\begin{equation}
h_{B} (\omega) G_{B}'' (\omega)
\end{equation}
and
\begin{equation}
h_{F} (\omega) G_{F}'' (\omega).
\end{equation}
These combinations are nothing more than the 
``Keldysh'' Green's
function well known from non-equilibrium physics\cite{rammer,kamenev}. In equilibrium, the
Keldysh Green's function satisfies the fluctuation dissipation theorem
\[
G_{K} (\omega) = [G_{R} (\omega) -G_{A} (\omega)]2 h (\omega) = - 4i G''
(\omega)h (\omega)
\]
where $G_{R}$ and $G_{A}$ are the retarded and advanced propagators,
respectively.
In other words, 
$h\cg'' (\omega) \equiv \frac{i}{4}G_{K}
(\omega)$. The point is, that we can formally imagine evaluating $Y$
in a general steady-state, replacing $G_{K}\rightarrow -4 i g'' (\omega)h(\omega)$
at the end of the calculation.  It is however, equally consistent to compute $Y$ in
equilibrium, and substitute $g'' (\omega)h (\omega)\rightarrow
g_{K}$. The functional $Y$ is closely related to the expectation value
of the interaction energy. If we associate an amplitude $\sqrt{\alpha }$
with the strength of each vertex, then the expectation value of the
interaction energy has the expansion
\[
\langle H_{I}\rangle = \sum \alpha ^{n} {Y}_{n} (\cg _{K},\cg _{R},\cg _{A})
\]
where $\cg_{K,R,A}$ are the fully renormalized non-equilibrium propagators.
 The functional  $Y$ is the weighted sum
\[
Y[\cg_{K},\cg_{R},\cg_{A}] = \sum \frac{\alpha ^{n}}{n} {Y}_{n} (\cg_{K},\cg_{R},\cg_{A})
\]
In this way, we can absorb all thermal functions into the Green's
functions, 
so that all temperature dependence is entirely contained within the
Keldysh Green's functions. To differentiate $Y$ with respect to
temperature, we need to determine $\delta Y/\delta \cg_{K}$. 
Unfortunately, $Y$ is not a true generating functional for the Keldysh
self-energies, and we must be very careful in taking the next step. 

Once we know the variation of $Y$ with
respect to the Keldysh Green's function, we can immediately compute the
variation with respect to temperature. 
Now, from (\ref{eq:deltay}), we see that we may write
\begin{eqnarray}\label{e}
\delta Y 
&=& \int \frac{d\omega}{ \pi  }
\str\left[ 
\ {\Sigma}' (\omega)
\hat  h (\omega)
\delta \cg'' (\omega)
+\hat  h (\omega){\Sigma}'' (\omega)
\delta \cg' (\omega)
\right] 
.
\end{eqnarray}
If we now identify 
\begin{eqnarray}\label{f}
\hat h (\omega)\cg '' &\equiv &\frac{i}{4}\cg_{K}\cr
\hat h (\omega)\Sigma '' &\equiv &\frac{i}{4}\Sigma_{K}
\end{eqnarray}
as the Keldysh components of the propagator and self-energy,
respectively, it is very tempting to write 
\begin{eqnarray}\label{g}
\delta Y 
= \int \frac{d\omega}{ \pi  }\str\left[ \left(\frac{i}{4} \right)
\ {\Sigma}' (\omega)
\delta \cg_{K} (\omega)+\left(\frac{i}{4} \right){\Sigma}_{K} (\omega)
\delta \cg' (\omega)
\right] 
.
\end{eqnarray}
This form would always be correct if $Y$ were a true generating functional
for Keldysh propagators. Unfortunately, in the Keldysh approach, the
distinction between the ``measurement'' and ``response'' vertices
means that in general, the differential of $Y$ with respect to
$\cg_{K}$ does not generate the real part of the self energy. 
Fortunately, this difficulty vanishes in the leading large $N$
approximation, which is sufficient to generate a wide class of
non-crossing approximation schemes. 
Diagrams involving vertex corrections do not satisfy
this relationship, and there are corrections to the above
diagram. Interestingly enough however, our large $N$ expansion does absorb
all the ``RPA'' diagrams of the interaction lines into explicit
propagators, enabling the entropy formula to be derived for interacting
collective modes and fermions.

With this provise, we write 
\[
\frac{\delta Y}{
\delta \cg_{K} (\omega)} = \frac{i}{4}\Sigma' (\omega)
\]
where $\Sigma ' = \frac{1}{2}\left( 
\Sigma (\omega-i\delta )+
\Sigma (\omega+i\delta )\right)$. 

It is worth making a short diversion at this point to explicitly
demonstrate that this relationship works in the large $N$ limit. 
Consider the leading 
order form for $Y$ in the large $N$ expansion, 
\begin{equation}
Y_{1}=  K\times N \times \left(\frac{1}{\sqrt{N}} \right)^{2}T^{2}\sum_{k,q}G_{c} (k)G_{\chi} (q-k)G_{b} (q)
\end{equation}
where $k\equiv (i\omega_{n}\vec{k})$, $q\equiv (i\nu_{n},\vec{q})$ is
a shortened space-time notation.  When we carry out the Matsubara sums
in this expression, 
we obtain
\begin{equation}
Y = - K \int_{-\infty}^\infty \frac{d\omega}{\pi}\frac{d\nu}{\pi}
\sum_{\vec k ,{\vec{q}}}
 \left\{
G_{c}' G_{\chi}'' G_{b}'' 
( h_{\chi} h_{b}- \frac{1}{4})
+ 
G_{c}'' G_{\chi}' G_{b}'' h_{c} h_{b}
+ 
G_{c}'' G_{\chi}'' G_{b}' h_{c} 
h_{\chi} 
 \right\},
\end{equation}
where we have used the short-hand notation $G_{c}\equiv G_{c} (k)$,
$h_{c}\equiv h_{c} (\omega)$, $G_{\chi }\equiv G_{\chi} (q-k)$ etc,
which can be rewritten as 
\begin{equation}
Y =  \frac{K}{4} \int_{-\infty}^\infty \frac{d\omega}{2\pi}\frac{d\nu}{2\pi}
\sum_{\vec k ,{\vec{q}}}
{\rm Re} \left\{ 
G_{Rc} G_{K\chi}
 G_{Kb}
+ 
G_{Kc}G_{R\chi} G_{Kb}
+ 
G_{Kc} G_{K\chi} G_{Rb} + G_{R\chi}G_{Rc}G_{Ab}
 \right\},
\end{equation}
This last expression can also be derived using Keldysh formalism to
calculate the leading order expression for $\langle H_{I}\rangle
$\cite{kamenev}. 
The variation of this expression with respect to $G_{K}$ gives the
leading order expressions for the boson, $\chi $ and conduction
self-energy, thus for instance,
\begin{eqnarray}\label{h}
\frac{\delta Y}{\frac{i}{4}\delta G_{Kb} (q)}&=& 
-\frac{i}{4}
k \sum_{\vk }\int \frac{d\omega_{c}}{\pi}\left[ G_{c}' (k) G_{K\chi} (q-k)
+ 
G_{Kc} (k) G_{\chi}' (q-k)
\right]\cr
&=& -
k \sum_{\vk }\int \frac{d\omega_{c}}{\pi}\left[ G_{c}' (k) G_{\chi} ''(q-k)h (q-k)
+ 
G_{c}'' (k)h (k) G_{\chi}' (q-k)
\right]
\end{eqnarray}
where the replacement $K\rightarrow k=\frac{K}{N}$ occurs because $Y$
contains the contribution of all $N$ spin channels of
$G_{Kb}$, and we are only differentiating with respect to one of
them.  This result can be independently confirmed using Matsubara techniques.

Let us now return from this digression and continue to calculate the
temperature dependence of $Y$. 
When we differentiate the general expression with respect to
the temperature, we only need to keep track of how each of the thermal
functions changes. 
This leads to the contribution, 
\begin{eqnarray}\label{i}
\hat  h (\omega)\cg '' (\omega)= \frac{i}{4}\cg_{K} (\omega)
&\rightarrow& \hat h (\omega)\cg ''
(\omega) + \frac{ d\hat  h (\omega)}{dT }\cg '' (\omega)\delta  T\cr
&=& \frac{i}{4}[\cg_{K} (\omega)+ \delta \cg_{K} (\omega)]
\end{eqnarray}
where
\begin{eqnarray}\label{eq:change}
\frac{i}{4}\delta \cg_{K} (\omega) = \frac{d \hat  h
(\omega)}{dT }\cg '' (\omega)\delta T
\end{eqnarray}
We can now combine these results to  obtain
\begin{eqnarray}\label{j}
\left.\frac{\partial Y}{\partial T} \right|_{\cg } 
&=& 
\int
\frac{d\omega}{\pi}
\str \left[\frac{\delta Y}{\frac{i}{4}\delta \cg_{K} (\omega)}\left(\frac{d \hat  h
(\omega)}{dT }\cg '' (\omega) \right) \right]\cr
&=& 
\int
\frac{d\omega}{\pi}
\str \left[\underline{\Sigma }' (\omega)\left(\frac{d \hat  h
(\omega)}{dT }\cg '' (\omega) \right) \right]
\end{eqnarray}
so that 
\begin{eqnarray}\label{k}
S_{2}=-\left.\frac{\partial Y}{\partial T} \right|_{\cg } = -\int
\frac{d\omega}{\pi} \str\left[
\left(
\frac{d\hat h (\omega)}{dT}
\right) {\rm Re}{\Sigma} (\omega)
\cg '' (\omega) \right] .
\end{eqnarray}
This is a key element in our proof of the entropy equation. 

When we add $S_{1}$ and $S_{2}$ together, $S_{2}$ partially cancels
the second-term in $S_{1}$, yielding the final answer
\begin{equation}\label{entropy2}
S = S_{1}
+ S_{2} =
\int \frac{d\omega}{\pi} \str \left[\left(\frac{d \hat h
(\omega)}{dT} \right)
\left({\rm Im} \ln \left[- \cg^{-1} (z ) \right]+ {\rm
Im}\Sigma (z) {\rm Re}\cg  (z)
 \right)_{z=\omega-i\delta }
 \right]
\end{equation}
We may replace $\hat  h\rightarrow \hat  n$ and $\str\rightarrow {\rm
Tr}_{B,F} $ to recover (\ref{entropy}).
There are a few important points to be made about this result
\begin{itemize}

\item Our ability to write the entropy in terms of the Green's
functions only works because in the leading order approximation for
$Y$, there are no hidden collective modes
which carry the entropy.  

\item This approach is general, and can, for example be applied to the
interacting electron gas, and various interacting plasmas, such as the
quark gluon plasma\cite{blaizot}.
By treating the interaction line as an independent particle
(photon), the leading order approximation generates a generalized RPA-Eliashberg
scheme for the self-consistent interaction and electron
self-energies. Note that the entropy formula  can not be used in an
RPA scheme in which the interaction line is not treated as an
independent particle\cite{pethick}, and it does not work if one
includes vertex corrections. 

\item It may be possible in future work to evaluate a more general
expression for $\delta Y/\delta \cg_{K} (\omega)$, permitting one to
generalize the entropy formula to higher order approximations for
$Y$. 

\end{itemize}

\section{Conservation laws and sum rules}

Sum rules are intimately related to the existence of conserved
charges. For example, in the Kondo lattice, there are three
independent types  of conserved charge: the electric charge (in each
channel), the total spin
and the conserved number of  bosons at each site. 
Each conserved charge 
corresponds to a gauge invariance of the Lagrangian.  For example, the
electric charge (for channel $\nu$) is
\begin{equation}
Q_{\nu} = \sum_{\vk, \al} c^{\dg}_{\vk \nu\al} 
c_{\vk \nu \al} 
\end{equation}
and this quantity is associated with the global gauge transformation
\begin{equation}
c_{\vk \nu \al} 
\rightarrow e^{i \theta_{\nu} (\tau )}c_{\vk \nu \al} , \qquad 
\chi _{j \nu } 
\rightarrow e^{-i \theta_{\nu} (\tau )}\chi _{j \nu } 
\end{equation}
The opposite sign in the exponents expresses the opposite charge of
the two fields. 
Similarly, the conserved number of bosons at each site
\begin{equation}
Q_{bj} = \sum_{\alpha }b\dg _{j\alpha}b _{j\alpha}
\end{equation}
generates the local  gauge transformation
\begin{equation}
b_{j\alpha }\rightarrow e^{i \theta_{b} (j,\tau )}b_{j\alpha }, \qquad 
\chi _{j \nu } 
\rightarrow e^{i \theta_{b}(j,\tau )}\chi _{j \nu } 
\end{equation}
In phases where these symmetries are unbroken, they give
rise to sum rules, which we now derive.

The general form of these gauge transformations is given by 
\begin{equation}
\psi_{\zeta} \rightarrow e^{ i \theta (\tau )q_{\zeta} } \psi_{\zeta}
\end{equation}
We can relate this gauge invariance to the
conserved charge by examining how the time-derivative terms in the
action behave under this transformation. 
In general, the time derivative part of the action has the form
\begin{equation}
S_{0}=\int_{0}^{\beta }
d \tau  \psi\dg \underline{\gamma}
\partial_{\tau}\psi
\end{equation}
where the elements $\gamma_{\zeta}$
of the diagonal matrix $\underline{\gamma}$
are unity, $\gamma_{\zeta}=1$ 
for conventional fermions or bosons and zero, $\gamma_{\zeta}=0$  for
Hubbard Stratonovich fields without any short-time dynamics, such as
the Grassmanian 
field $\chi_{j\mu}$ introduced in the Kondo lattice.
We can relate $\underline{\gamma}$ to the frequency
dependence of ${\cg}_{0}^{-1}$,
\begin{equation}\label{theidentity}
\underline{\gamma} = \frac{d {\cg}_{0}^{-1} (\omega)}{d \omega}.
\end{equation}

Under a time-dependent gauge
transformation, the change in the action is given by
\begin{eqnarray}\label{l}
\Delta S&= &
\left(\int_{0}^{\beta } d \tau  
\psi\dg e^{- i \theta
\hat q }
\underline{\gamma}
\partial_{\tau}e^{i \theta \hat  q }\psi 
 \right)-S_{0}\cr
&=&  i \int_{0}^{\beta } d
\tau  
\partial_{\tau }\theta (\tau ) Q (\tau )\cr
&=& - i \int_{0}^{\beta } d
\tau  
\theta (\tau ) \partial_{\tau }Q (\tau )
\end{eqnarray}
where we have commuted the diagonal matrices $\hat q$ and $\underline{\gamma}$ to obtain
\begin{equation}\label{thecharge}
Q = \psi \dg (
\hat{q}\underline{\gamma}  ) \psi.
\end{equation}
Invariance of the action under time-dependent gauge transformations,
implies that $\partial_{\tau}Q (\tau)=0$, i.e. the
charge is conserved.
In this way, we see that 
$\hat q \underline{\gamma}$ is the single particle operator associated
with the charge $Q$. Notice that the full charge operator
only depends on those fields with a time derivative. 
When the system is probed at short times, only these fields carry the charge.
However, at long times, it is the gauge charge $\hat q$ which determines the
charge of physical low energy excitations.

Using 
(\ref{thecharge}), we can now write the 
expectation value of the 
the charge $Q$ in the following compact form
%\begin{eqnarray}
\begin{equation}
\langle \hat{Q} \rangle = 
\langle 
 \psi^{\dg}_{\al} 
\hat{q} \underline{\gamma}  \psi_{\al} \rangle
= - \frac{1}{\be} \Str  \left[
 \hat q  \underline{\gamma} \cg\right],
\end{equation}
At zero temperature, provided there {\sl is } a scale to the
excitations, 
then the summations over discrete Matsubara frequencies 
can be replaced by a continuous integral. In particular, 
\begin{equation}
T \Str [A] \rightarrow
\int_{-i\infty}^{i\infty}\frac{d\omega}{2 \pi i } \str[
A (\omega)]e^{\omega 0^{+}}
\end{equation}
where $\str[A]$ without an underline indicates a supertrace purely
over the spatial variables of A. 
This enables us to write
\begin{equation}
\label{eq:conserved-Q1}
\langle \hat{Q} \rangle =   
- \int_{-i \infty}^{i \infty} \frac{d \omega}{2 \pi i} 
\str \left[ \hat q  \underline{\gamma}
\cg (\omega) \right],
\end{equation}
where for clarity, we have temporarily suppressed the convergence
factor $e^{\omega 0^{+}}$.
Using (\ref{theidentity}), 
$\underline{\gamma}  = (d {\cg}_0^{-1})/(d \omega)$, we then obtain
at $T=0$
\begin{equation}
\label{eq:conserved-Q}
\langle \hat{Q} \rangle =   
- \int_{-i \infty}^{i \infty} \frac{d \omega}{2 \pi i} 
\str \left[ \hat q  
\frac{d {\cg}_0^{-1}}{d \omega}\cg \right].
\end{equation}

Now let us consider the effect of the gauge transformation on the Free energy.
If we carry out a gauge transformation
$\psi_{\zeta} \rightarrow e^{i \theta (\tau )q_{\zeta} }\psi_{\zeta}$,
with $\theta (\tau )= \Delta \omega \tau $, then diagrammatically, this
has the effect of shifting the frequencies associated with each propagator,
$\omega_{\zeta}\rightarrow \omega_{\zeta} + q_{\zeta} \Delta
\omega$, where $q_{\zeta}$ is the gauge charge of the propagating particle. In the zero temperature limit,
this shift in frequency can be made infinitesimally small, $\Delta \omega_{\zeta}\rightarrow \delta \omega_{\zeta}$.

Diagrammatically,  the conservation of charges corresponds to the
existence of closed loops (Fig. \ref{fig6}.) in the skeleton digrams of $Y$ through which
the conserved charge circulates.   The gauge
transformation causes the frequency variable in each of these closed loops
to shift by an amount $q \Delta \omega$. This change does not affect
the value of $Y$, and vanishing of the change in $Y$ then gives rise to
new Ward identities.

\figwidth=7cm
\fg{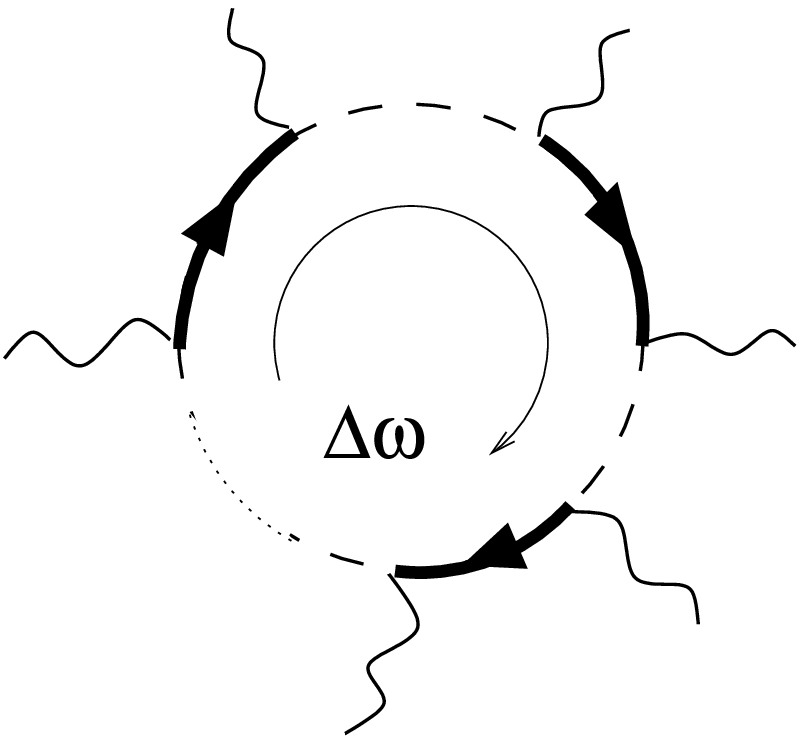}{fig6}{Illustrating the shift in frequencies around a
closed loop in $Y[\cg ]$ for the Kondo lattice, generated by the gauge transformation
associated with charge conservation.}
\figwidth=14cm

When the frequency around the closed loops is shifted, each
propagator entering into $Y$ changes, with $\delta {\cg}_{\zeta} (\omega)
=
\frac{d
\cg}{d \omega} q_{\zeta} \delta \omega$, so that the change
in $Y$ at zero temperature is 
\begin{eqnarray}\label{m}
\delta Y &=&- \int_{-i\infty}^{i\infty}\frac{d\omega}{2 \pi i } \str[
\Sigma (\omega) \delta \cg (\omega)]\cr
&=&- \int_{-i\infty}^{i\infty}\frac{d\omega}{2 \pi i } \str[
\Sigma (\omega)
 \frac{d \cg}{d \omega} 
\hat  q]\delta \omega = 0,
\end{eqnarray}
so that 
\begin{eqnarray}
0 = \frac{\delta Y[\cg]}{\delta \omega}
&=& 
\int_{-i \infty}^{i \infty} \frac{d \omega}{2 \pi i}
\str \left[ \Sigma 
 \frac{d \cg}{d \omega} 
\hat q
\right].
\nonumber 
\end{eqnarray}
Integrating by parts, we obtain the Ward Identity
\begin{equation}\label{eq:stationarity}
 \int_{-i \infty}^{i \infty} \frac{d \omega}{2 \pi i}
\str \left[\frac{d \Sigma}{d \omega} \cg \hat q
\right]=0.
\end{equation}
Combining eqs.~\eqref{eq:conserved-Q} and \eqref{eq:stationarity} we obtain
\begin{eqnarray}
\label{eq:ward-identity}
\langle \hat{Q} \rangle 
&=&
 - \int_{-i \infty}^{i \infty} \frac{d \omega}{2 \pi i} 
\str \left[ \hat q 
\frac{d}{d \omega} 
\overbrace {\left( {\cg}_0^{-1} - \Sigma \right)}^{\cg^{-1} (\omega)}
 \cg \right]e^{\omega 0^{+}}
\nonumber \\
&=& - \int_{-i \infty}^{i \infty} \frac{d \omega}{2 \pi i}
\frac{d}{d \omega} 
\left(\str\left[ \hat q 
\ln(- \cg^{-1}) \right]
 \right)e^{\omega 0^{+}}
,
\end{eqnarray}
where we have explicitly reinstated the convergence factor. 
Folding the integration contour around the negative real axis, (as
dictated by the convergence factor) we then obtain
\begin{eqnarray}\label{n}
\langle \hat{Q} \rangle &=&
- \int_{- \infty}^{0} \frac{d \omega}{\pi}
\frac{d}{d \omega} \frac{1}{{2i}}\left (
{\str\left[ \hat q 
\ln(- \cg^{-1} (\omega-i\delta )) \right]
- 
\str\left[ \hat q \ln(- \cg^{-1} (\omega+i\delta )) 
\right]
}
 \right)
\cr
&=&- \int_{- \infty}^{0} \frac{d \omega}{\pi}
\frac{d}{d \omega} 
{\rm Im}\ \str\left[ \hat q 
\ln(- \cg^{-1} (\omega-i\delta )) \right]\cr
&=&
- \left[ 
\frac{1}{\pi}
{\rm Im}\ \str\left[ \hat q 
\ln(- \cg^{-1} (\omega-i\delta )) \right]\right]_{-\infty}^{0}.
\end{eqnarray}
The presence of the minus sign inside
the logarithms here is chosen so that the lower bound of the integral vanishes.
We are then able to write
\begin{eqnarray}\label{eq:withbosons}
\langle \hat{Q} \rangle 
= - \left. 
\frac{1}{\pi}
{\rm Im}\ \str\left[ \hat q 
\ln(- \cg^{-1} (\omega-i\delta )) \right]\right|_{\omega=0}.
\end{eqnarray}
As it stands, the trace in this  expression involves both bosonic
and fermionic components.  
The latter will give rise to
Fermi surface volume contributions in the lattice, and to
phase shift terms in the case of impurity models. What about the
bosons?
General arguments lead us to believe that the bosonic
component in the trace vanishes. The important point here,
is that the right-hand side of this sum 
counts the number of one-particle states that drop to {\sl negative energy}.
Whereas fermions can acquire a negative energy through the formation
of a resonance or Fermi surface, bosons can not - they only condense.
If the bosons condense, then the symmetry they are associated with is
broken, and we can no longer apply the sum rule. If they do not
condense, then we expect the associated phase shifts to be zero. 
The
final form of the sum rule is then
\begin{eqnarray}\label{o}
\langle \hat{Q} \rangle 
=
\frac{1}{\pi}
{\rm Im}\ \tr\Biggl[  \hat q 
\ln(- {\cg}_{F}^{-1}) \Biggl]_{\omega= 0-i\delta },
\end{eqnarray}
where for clarity, we have suppressed the frequency argument of the
Green's function.

\section{Kondo impurity model}

As a first illustration of the sum rules, let us apply them to the
Kondo impurity model.  
There are two sum rules to consider here
corresponding to conservation of $Q_{c}$ and $n_{b}= 2S = K$
in the case of the perfectly screened solution. Let us first consider
\begin{equation}
n_{b}= \sum_{\alpha} b\dg _{\alpha}b _{\alpha}.
\end{equation}
Even though the $\chi $ fermions do not enter into this conserved
charge, they carry the gauge charge, and we have  $(q_{c},q_{b},q_{\chi })= (0, 1, 1)$.
Applying the sum rule (\ref{eq:withbosons}), we deduce 
\begin{eqnarray}\label{p}
\langle n_{b}\rangle  
= - 
\frac{1}{\pi}
{\rm Im}\ \tr\biggl[ 
\ln(- {\cg}_{b}^{-1} ) \biggr]_{\omega=i\delta }
+
\frac{1}{\pi}
{\rm Im}\ \tr\biggl[ 
\ln(- {\cg}_{\chi }^{-1} ) \biggr]_{\omega=i\delta },
\end{eqnarray}
where for the moment, we have retained the bosonic trace.
When we carry out the trace over the spin components of the
boson and the channel components of the $\chi $ fermion, we obtain
\begin{equation}
\langle n_{b}\rangle = -N \frac{\delta_{b}}{\pi} + K \frac{\delta_{\chi}}{\pi},
\end{equation}
where
\begin{eqnarray}\label{q}
\delta_{b} &=& 
{\rm Im}\ 
\left[\ln(\lambda + \Sigma_{b} (0-i \delta )+i\delta )  \right],\cr
\delta_{\chi } &=& 
{\rm Im}\ 
\left[\ln(\frac{1}{J_{K}} + \Sigma_{\chi } (0-i \delta)) 
 \right],\end{eqnarray}
are the ``phase shifts'' associated with the boson and holon fields. Note in passing that at zero temperature, $\Sigma_{b}(0-i \delta)$ and $\Sigma_{\chi}(0-i\delta)$ are purely real.
The argument of the logarithm in $\delta_{b}$, 
is the renormalized
chemical potential $\lambda^{*}= \lambda + \Sigma_{b} (0)$ 
of the boson field.  In the ``filled shell'' configuration
$2S=K$, the formation of a perfectly screened Kondo singlet
generates a gap for the addition of extra Schwinger bosons to the
ground-state, which forces
$\lambda^{*}>0$,  and $\delta_{b}=0$ as argued previously.  
It follows that
\begin{equation}
\langle n_{b}\rangle = K \frac{\delta_{\chi}}{\pi}.
\end{equation}
But  $n_{b}=K$ is part of the constraint, so it follows that 
$\delta_{\chi } = \pi$.  This result can be understood 
by relating  $\delta_{\chi }$
to the sign of the renormalized Kondo coupling constant, as follows
\begin{equation}
\frac{\delta_{\chi }}{\pi} = \frac{1}{\pi}
{\rm Im}\ \left[ 
\ln(\frac{1}{J_{K}} + \Sigma_{\chi } (0-i\delta )) \right] =
\theta (- {J_{K}^{*}}).
\end{equation}
In this way, we see that $\delta_{\chi }=\pi$ 
corresponds to a residual
``ferromagnetic'' coupling. 
This is consistent with our expectations, for 
when an additional Schwinger boson is added to the Fermi liquid, it
increases the 
impurity  spin by one half unit to form an underscreened Kondo
model, where the residual spin coupling is indeed ferromagnetic.  By contrast,
in an antiferromagnet or spin liquid, the local moments spins pair-condense, mutually
screening one-another. In this situation, an additional Schwinger boson
at any one-site, will now be free to undergo a Kondo effect with the
conduction electrons, so the residual spin-coupling is
antiferromagnetic. 
These
results are also confirmed in the large $N$ limit.

Let us now turn to the sum rule derived from conservation of
electron charge. Here the conserved quantity is
\begin{equation}
Q_{e} = \sum_{k,\nu,\sigma} c\dg_{\vk,\nu,\sigma}c_{\vk,\nu,\sigma}
\end{equation}
and the corresponding gauge charges are
$(q_{c},q_{b},q_{\chi }) = (1, 0 , -1)$, so the 
sum rule becomes
\begin{equation}\label{cond}
Q_{e} = \frac{1}{\pi}\left. \left( 
{\rm Im}\ \tr\left[ 
\ln(- {\cg}_{c}^{-1} )
 \right]
- \frac{1}{\pi}
{\rm Im}\ \tr\left[ 
\ln(- {\cg}_{\chi }^{-1} ) \right]\right)\right|_{\omega= 0 - i \delta }
\end{equation}
Now in an impurity model, it is convenient to subtract off the total charge
in the absence of the impurity, given by 
\begin{equation}
Q_{e}^{(0)} = 
\frac{1}{\pi}
{\rm Im}\ \tr\biggl[ 
\ln(- {\cg}_{c0}^{-1}) \biggr]
_{\omega= 0 - i \delta },
\end{equation}
where ${\cg}_{c0}\equiv {\cg}_{c}^{(0)}$ is the non-interacting propagator of the
conduction electrons.  If we subtract this from (\ref{cond}),
we can combine the conduction electron traces 
and replace $\ln(- {\cg}_{c}^{-1} )- \ln(- {\cg}_{c0}^{-1} ) 
= \ln( {\cg}_{c0} {\cg}_{c}^{-1} )= \ln (1 - {\cg}_{c0}\Sigma_{c}) 
$. The change in total charge is then given by
\begin{eqnarray}\label{r}
Q_{e} -Q_{e}^{(0)}=\frac{1}{\pi}
{\rm Im}\ \tr\biggl[ 
\ln( 1 - {\cg}_{c0} \Sigma_{c}) \biggr]_{\omega=-i\delta }
- \frac{1}{\pi}
{\rm Im}\ \tr\biggl[ 
\ln(- {\cg}_{\chi }^{-1}) \biggr]_{\omega=-i\delta }.
\end{eqnarray}
According to the Anderson-Clogston compensation theorem, 
the total change in electron charge due to a Kondo or Anderson impurity
vanishes in the infinite band-width limit, and at finite band-width
the change is of order the ratio of the Kondo temperature to the
bandwidth $\Delta Q_{c}= O (T_{K}/D)$, and can be neglected\cite{clogston61,compensationnote}.
Setting $\Delta Q_{e} =Q_{e}-Q_{e}^{0}= 0$,
we obtain
\begin{equation}
\Delta Q_{e}= 0 = N K \frac{\delta_{c}}{\pi} - K \frac{\delta_{\chi}}{\pi}
\end{equation}
where we have identified
\begin{equation}\label{phasec}
\delta_{c} = {\rm Im}\ \tr_{\vk }\biggl[ 
\ln( 1 - {\cg}_{c0} \Sigma_{c}) \biggr]_{\omega=-i\delta }
\end{equation}
where the trace is purely over momenta. 
If we expand the logarithm using a power series,
we see that the trace over internal momenta in (\ref{phasec}) is accomplished by
replacing the momentum dependent Green's function ${\cg}_{c}^{(0)} (\vk,\omega)$ by the local electron
Green's function $g_{0} (\omega) = \sum_{\vk }{\cg}_{c}^{0} (\vk,\omega)$, so that 
\begin{equation}
\delta_{c} = {\rm Im} \ln  \left[1 -
g_{0} (\omega)\Sigma (\omega) \right]\biggr\vert_{\omega= 0 - i\delta }.
\end{equation}
In the large band-width limit, $g_{0} (\omega-i\delta)\rightarrow 
i \pi \rho $, where $\rho $ is the density of states per spin per
channel,  so that 
\begin{equation}
\delta_{c} = {\rm Im} \ln  \left[1 -
i \pi \rho \Sigma (0) \right]
\end{equation}

There is thus a direct link between the phase shift of the $\chi $
fermion and that of the conduction electrons. Combining this with the
earlier result $\delta_{\chi}= \pi$, we obtain
\begin{equation}
\delta_{c} = \frac{\delta_{\chi }}{N} = \frac{\pi}{N}
\end{equation}
Notice that
\begin{itemize}
\item in the large $N$ limit
the phase shift identity permits
us to relate the conduction electron phase shift $\delta_{c}$, which is $O (1/N)$
to the $\chi $ phase shift
$\delta_{\chi }$, which is finite in the large $N$ limit.
In this way, the sum rules enable us to study the phase shift and Fermi surface
volume changes of the Kondo impurity and lattice, in the large $N$
limit. 

\item the development of the conduction electron phase shift does not
occur because conduction electron states drop beneath the Fermi sea:
it is a consequence of the injection of new quasiparticle states into
the Fermi sea as a response to the formation of spinless, charged holons. 

\end{itemize}

\section{Sum rule for the Kondo lattice}

When we come to consider the Kondo lattice,
there are a number of additional 
subtleties that must be entertained. One of the most important aspects
of the discussion here concerns the possibility that the extremal solutions
to $F[\cg]$ break the $U (1)$ symmetry of the bosons, to produce 
``Higg's'' phase
in which the Schwinger bosons either pair condense, or
hybridize between sites. In the large $N$ solutions to the model,
these phases appear to develop as a precursor to the formation of
the antiferromagnet, where the bosons themselves condense.
The main effect on the sum rules is two fold:
\begin{itemize}

\item  we can
not assume that the $\chi $ fermion is localized. This implies that we
must consider the momentum dependence of the $\chi $ fermion.

\item once the Schwinger bosons pair condense, the 
Ward identities associated with the conservation of $n_{b}$ no longer apply,
because the Luttinger Ward functional $Y[\cg]$ is no longer invariant
under shifts of the boson frequency. 

\end{itemize}

However, we can still take advantage of the Ward identity associated
with charge conservation. Provided that the Schwinger boson is
uncondensed, i.e, there is no magnetism, then charge conservation
guarantees that
\begin{equation}
 \int_{-i \infty}^{i \infty} \frac{d \omega}{2 \pi i}
\tr_{F} \left[\frac{d \Sigma}{d \omega} \cg \hat q
\right] = 0, \qquad \qquad (q_{c}=1, q_{\chi }=-1)
\end{equation}
Now in the paramagnet, the holon and conduction electron fields are
unmixed,
so we can replace
\[
\tr_{F}\rightarrow\tr_c + \tr_{\chi }.
\]
This is a special property of the paramagnet. 
By contrast, in the antiferromagnet, 
the Schwinger bosons are condensed, so the 
holon and conduction electron fields admix, and both  $\Sigma $ and
$\cg $ contain off-diagonal terms so that $\tr_F$ remains a single,
integral trace over the two admixed fields.

For the paramagnet, we can write
\begin{eqnarray}\label{s}
0&=&- \int_{-i \infty}^{i \infty} \frac{d \omega}{2 \pi i}
\str \left[\frac{d \Sigma}{d \omega} \cg \hat q
\right], \qquad \qquad (q_{c}=1, q_{\chi }=-1)\cr
&=&
 \int_{-i \infty}^{i \infty} \frac{d \omega}{2 \pi i}
\tr \left[\frac{d \Sigma_{c}}{d \omega} {\cg}_{c} 
\right]
-  \int_{-i \infty}^{i \infty} \frac{d \omega}{2 \pi i}
\tr \left[\frac{d \Sigma_{\chi }}{d \omega} {\cg}_{\chi } 
\right]
\cr
&=&
 K \int_{-i \infty}^{i \infty} \frac{d \omega}{2 \pi i}
\int
\frac{d^{D}k}{(2 \pi)^{D}}
\left[N\frac{d \Sigma_{c} (\vk ,\omega)}{d \omega} {\cg}_{c} (\vk ,\omega) 
-
\frac{d \Sigma_{\chi } (\vk ,\omega)}{d \omega} {\cg}_{\chi } (\vk ,\omega) 
\right].
\end{eqnarray}
It is this identity that permits us to extend the Luttinger sum
rule to the case of the paramagnetic Kondo lattice. In the
paramagnetic Kondo lattice, the charge sum rule
\begin{equation}\label{cond2}
Q_{e} = \frac{1}{\pi}\left. \left( 
{\rm Im}\ \tr\left[ 
\ln(- {\cg}_{c}^{-1} )
 \right]
-
{\rm Im}\ \tr\left[ 
\ln(- {\cg}_{\chi }^{-1} ) \right]\right)\right|_{\omega= 0 - i \delta }
\end{equation}
now involves a trace over momentum:
\begin{equation}
\frac{Q_{e}}{K}= N \sum_{\vk }\frac{1}{\pi}{\rm Im}\ln \bigl[ \epsilon_{\vk
}+ \Sigma_{c} (\vk ,0-i \delta)+i\delta 
\bigr ]- 
\sum_{\vk }\frac{1}{\pi}{\rm Im}\ln \bigl[ 
\frac{1}{J_{K}}
+ \Sigma_{\chi } (\vk ,0-i \delta) 
\bigr ].
\end{equation}
The first term in this expression is the electron Fermi
surface volume,
\begin{equation}
\sum_{\vk }\frac{1}{\pi}{\rm Im}\ln \bigl[ \epsilon_{\vk
}+ \Sigma_{c} (\vk ,0-i \delta)+i\delta 
\bigr ]= \sum_{\vk }\theta (-E_{\vk})=
\frac{{\rm v}_{FS}}{(2\pi)^{D}} ,
\end{equation}
where the region where 
$E_{\vk }= \epsilon_{\vk }+ {\rm Re}\Sigma_{c} (\vk ,E_{\vk })$ is negative
defines the interior of the Fermi surface.
The second term can be interpreted in a similar way -
the momentum trace over the logarithm of ${\cg}_{\chi }^{-1}$
\begin{equation}
\sum_{\vk }\frac{1}{\pi}{\rm Im}\ln \bigl[ 
\frac{1}{J_{K}}
+ \Sigma_{\chi } (\vk ,0-i \delta)
\bigr ]= \sum_{\vk}\theta [-J_{K}^{*} (\vk )]
= 
\frac{{\rm v}_{\chi }}{(2\pi)^{D}} 
\end{equation}
can be seen as the volume of the
region in momentum space where the effective interaction
${J_{K}^{*} (\vk )}= - {\cg}_{\chi} (\vk ,0)^{-1}$
is negative, or ferromagnetic.
Following our earlier discussion, the ferromagnetic sign of the
residual interaction is a consequence of the fact that additional
spins
added to the state completely decouple from the Fermi sea. 
In the simplest scenario, 
$J_{K}^{*} (\vk )<0$ for all $\vk $, in which case
$\frac{{\rm v}_{\chi }}{(2\pi)^{D}} =1
$, and the sum rule becomes
\begin{equation}
n_{e}=\frac{Q_{e}}{K} =  N\frac{{\rm v}_{FS}}{(2\pi)^{D}} 
- 1
\end{equation}
where $n_{e}$ is the electron density per unit cell, per conduction
electron channel. 
It follows that the total Fermi surface volume expands by one unit per unit cell,
\begin{equation}
 N\frac{{\rm v}_{FS}}{(2\pi)^{D}} = n_{e}+1 
\end{equation}

Now if there is some region of momentum space where $J_{K}^{*}
(\vec{k})$ is not ferromagnetic, it follows that the $\chi $ fermions
will have a Fermi surface, and the excitation spectrum will now involve
charged, spinless fermions. In this phase, 
\begin{equation}
n_{e}=\frac{Q_{e}}{K} =  N\frac{{\rm v}^{*}_{FS}}{(2\pi)^{D}} 
- \frac{{\rm v}_{\chi }}{(2\pi)^{D}} 
\end{equation}
Such a ``spin-charge decoupled'' phase
could not develop from a Fermi liquid without a phase
transition. So long as the heavy electron fluid does not undergo
such a phase transition, the Fermi surface must remain large.

Let us now consider what happens in the antiferromagnet. In the
general higher dimensional case, 
the boson field condenses, causing 
the conduction and $\chi$ fields to hybridize to produce a single
species.  
In the antiferromagnet, it is more logical to make a
particle-hole transformation of the $\chi $ field, writing $\psi_{\vk
} = \chi \dg _{-\vk}$. In this case, $q_{\psi }= q_{c}$ have
the same charge, and the sum rule becomes simply
\begin{equation}
Q_{e} = \frac{1}{\pi}{\rm Im}Tr_{F}\left[\ln (-{\cg}_{F}^{-1})
\right]\biggr\vert_{\omega= - i\delta }
\end{equation}
where ${\cg}_{F}$ is the admixed propagator for the combined conduction
and $\psi$ fields. The right-hand side can not be separated into $\chi $ and conduction parts, and as such, 
defines  an admixed
set of Fermi surfaces,  with  an average Fermi surface volume 
which counts the total charge per unit cell
\begin{equation}
n_{e} = N \frac{\langle {\rm v}_{AFM}\rangle }{(2\pi)^{D}}.
\end{equation}
The Fermi surface volume for the antiferromagnet and the
paramagnet must then differ by $1/N$ th of a unit cell.  In the
special case of $N=2$, if the spatial unit cell doubles, this formal change
in Fermi surface volume is indistinguishable from the new size of the
Brillouin zone. However, even in this case, we can imagine more
general classes of antiferromagnet, such as an incommensurate helimagnet, where the
unit cell size is unchanged, but the total Fermi surface volume must jump.

\section{Spin-charge separation and quantum criticality.}

Although the sum rules do not provide us with any details of the
dynamics, they  provide stringent constraints on the way the 
spectrum of excitations can evolve in the Kondo lattice as we approach
the magnetic quantum critical point. 

One of the fascinating aspects of our sum rule
\begin{equation}
n_{e}= N \frac{{\rm v}_{FS}}{(2\pi)^{D}} - \frac{\rm v_{\chi }}{(2 \pi)^{D}}
\end{equation}
is that it suggests that a  Kondo lattice may develop low-lying, spinless
charged fermionic excitations near a quantum critical point. This is
clearly a controversial idea, the consequences of which we now explore in this discussion.
If we take this idea seriously, then two important points seem to emerge:

\begin{itemize}
\item In the heavy fermi liquid, spinons have ``ionized'' into
electrons and a background filled sea of holons.
\begin{equation}
b_{\sigma} \leftrightarrow e_{\sigma }^{-}  + \chi ^{+}
\end{equation}
It is this ionization process that lies behind the expansion of the
Fermi surface and the sum rule. 

\item There is an intimate link between the Kondo interaction 
and the formation of holons. The link between the  propagator of the holons
and the renormalized Kondo interaction
\begin{equation}
- {\cg}_{\chi} (\vq ,\omega) = \frac{1}{\frac{1}{J_{K}}+\Sigma_{\chi} (\vq
,\omega)} = J^{*}_{K } (\vq ,\omega)
\end{equation}
implies that poles in the holon spectrum correspond 
to divergences in the Kondo interaction. As magnetism drives the Kondo
interaction back to weak coupling,  we expect $J^{*} (\vq ,\omega)$
to pass through infinity, so the holons should become gapless at
some point in this process. 
\end{itemize}

We can in fact use the Luttinger Ward approach to gain insight into the
low-energy thermodynamics of the Kondo lattice\cite{luttinger3}. 
Provided the replacement of frequency
sums by continuous integrals is valid at low temperatures, Luttinger
argued that the low energy thermodyanmics is determined 
by the leading logarithm in the Free energy functional (\ref{notsobigdeal}), evaluated with
the zero-temperature Green's function. In our case, this leads to the relation
\[
F (T)\approx \sum T{\Str}\left[\ln  \left(- \cg_{(0)}^{-1} \right)\right]
,\qquad \qquad (T\rightarrow 0)
\]
where $\cg_{( 0)}$ is the zero temperature Green's function. Carrying out
the frequency integrals, this implies that the low temperature entropy
is given by
\begin{eqnarray}\label{t}
S (T)&=&  \int \frac{d^{D}k}{(2\pi)^{D}}\int \frac{d\omega}{\pi}
\overbrace {\left(
 \frac{dn}{dT} {\rm Im} \tr \ln - \cg_{b}^{-1} (\vk ,\omega)+
 \frac{df}{dT} {\rm Im} \tr \ln - \cg_{\chi }^{-1} (\vk ,\omega)
 \right)}^{\hbox{spinon and holon entropy}}
\cr
&+&\int \frac{d^{D}k}{(2\pi)^{D}}\int \frac{d\omega}{\pi}
\overbrace {\left( \frac{df}{dT} {\rm Im} \tr \ln - \cg_{c}^{-1} (\vk ,\omega)
 \right)}^{\hbox{Fermi liquid}}
\end{eqnarray}
where the zero temperature propagators are to be used.
The last term in this expression provides the $T-$ linear entropy of the Fermi liquid,
the first two terms are the ``spinon'' and ``holon'' contributions.
For a Landau  Fermi liquid to form, these last terms must clearly be
gapped at low temperatures. This feature is observed in the large
$N$ solutions\cite[]{rech}.  
What it also makes clear however, is that the holons can
not become gapless without a thermodynamic departure from Fermi liquid
behavior, which 
implies some kind of quantum phase
transition.

However, even though the holons are gapped in the Fermi liquid, 
they can exist as low-lying spinless, charged excitations. 
It is particularly interesting to speculate that
the gap to holon formation closes at a quantum critical point.
This is a possibility that can be studied in the large $N$ limit,
but the general arguments should hold at finite $N$, 
so long as the Luttinger Ward approach is valid.
One way to go beyond the large $N$ limit, is to
to carry out a fully self-consistent treatment of the leading
Luttinger Ward Free energy functional, updating the conduction
electron self-energy  and feeding the full conduction electron Green's
functions into the calculation of the self energy for the $\chi$ and
Schwinger boson field. Such an approximation goes selectively beyond
the large $N$ limit, but it will satisfy the Ward
identities. Moreover, assuming that  the ``filled shell'' stability of the
fully screened ground-state is not an artifact of the large $N$ limit,
we also expect the gap in the Schwinger boson and the $\chi $
fermion particle spectrum to be preserved at finite $N$. 

%%From the link between the Kondo interaction and the holon propagators,
%%qualitatively, we expect that as 
%%as magnetism suppresses the Kondo effect, the Kondo interaction will
%%be driven upwards, 
%%driving the holon energy band towards the Fermi surface. At some
%%point, the holon spectrum may become gapless. Two possibilities seem to emerge 
%%\begin{enumerate}
%%
%%\item  the holon spectrum first touches the Fermi surface at a few
%%high-symmetry points in the Brillouin zone. 
%% 
%%\item alternatively, that - since spinons become almost gapless at points
%%in momentum space as antiferromagnetic correlation length grows, 
%%the momenta of low-energy holons will correspond to the 
%%the electron Fermi surface, shifted by the momentum of the condensing spinons.
%%This would give rise to a ``ridge'' of low
%%energy holons that shadows the  heavy electron Fermi surface, ultimately
%%annihilating with it. 
%%
%%\end{enumerate}

Another way to check the picture emerging from this approach would be to examinelow-dimensional systems.  Various authors have examined the
possibility of a transition from small to large Fermi surface in
the one-dimensional spin-1/2 Kondo lattice\cite[]{miranda,pivivarov}.  The difficulty with
this model, is that the gaplessness of the Heisenberg spin-1/2
chain tends to make the Kondo coupling relevant, no matter how
small the Kondo temperature.
A model that avoids these difficulties is the $S=1$, two channel
Kondo lattice. This model involves a Haldane spin-1 chain coupled via
Kondo interactions to two one dimensional, non-interacting
Hubbard models, as shown in Fig. \ref{fig7}. 
\figwidth=12cm
\fg{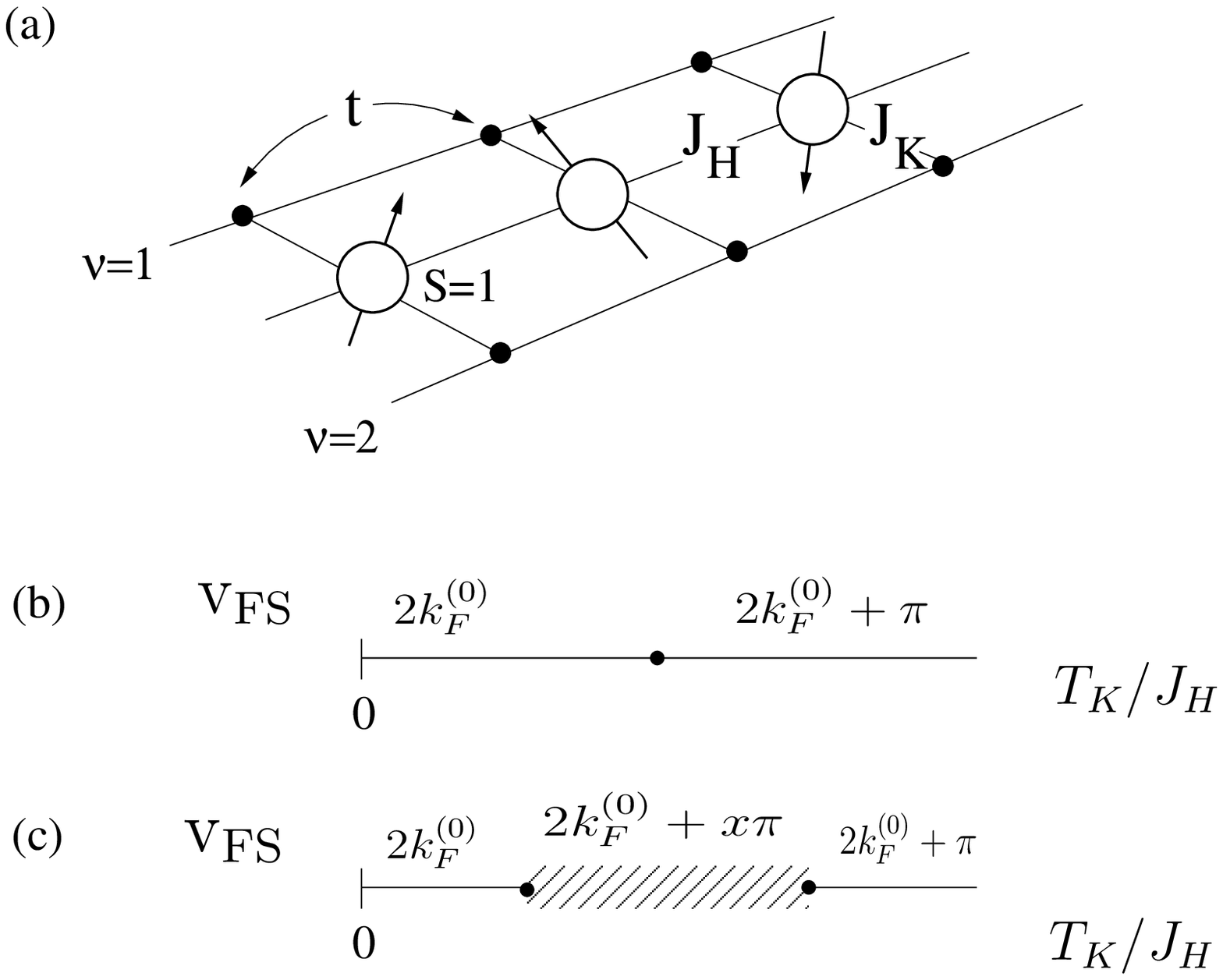}{fig7}{(a) The two-channel, spin $S=1$ Kondo
chain. (b) In the simplest scenario, the transition from a Haldane chain
with small Fermi surface, to the heavy electron phase with large Fermi
surface will occur via a single phase transition. (c) The sum rules allow
for an intermediate phase with a Fermi surface that lies between these
two extremes.
}

When the Kondo interaction is weak, the Haldane chain is gapped,
and the ``Fermi surface'' of the Kondo lattice is small, giving rise
to Friedel oscillations  at wavevector $Q= 2k_{F}^{(0)}$.
When the
Kondo interaction is strong, a Kondo lattice with a large Fermi
surface is expected to form. If the ``large'' and ``small'' 
Fermi surface phases, where $Q= 2 k_{F}^{(0)}+ \pi$. The way these two
phases are linked is particularly interesting. If they are linked by a
single quantum critical point, then we expect $Q$ to simply jump
at this point (Fig. \ref{fig7}b.). However, if the holons form a gapless
phase then over an intermediate range  of parameters, we would expect
the Friedel oscillations to exhibit an intermediate wavevector (Fig. \ref{fig7}c.), 
\begin{equation}
Q=  2k^{0}_{F} + x \pi.
\end{equation}

\vspace{40pt}

We are grateful to Mathew Fisher, Kevin Ingersent, Andreas Ludwig,
Catherine P\' epin and Gergely Zarand for discussions related to
this work. 
This research was supported by the
National Science Foundation grant NSF DMR 0312495 and PHY99-07949. The authors would like to thank
the hospitality of the KITP, where most of this research was carried
out.

%-------------------


\begin{thebibliography}{99}
%-------------------

\bibitem{luttinger}J. M. Luttinger, Phys. Rev. {\bf 119},  1153 ( 1960).

\bibitem{luttinger2}J. M. Luttinger and J. C.  Ward, Phys. Rev. {\bf 118},
1417 (1960).

\bibitem{luttinger3}J. M. Luttinger, Phys. Rev. {\bf 121}, 1251-1258 (1961).

\bibitem{friedel} J. Friedel, Phil. Mag. {\bf 43}, 153 (1952). 

\bibitem{langer} J. S. Langer and V. Ambegaokar, Phys. Rev. {\bf 121}, 1090 (1961) .

\bibitem{langreth}D. C. Langreth, Phys. Rev.  {\bf 150}, 516 (1966).

\bibitem{abrikosov}A. A. Abrikosov, Physics {\bf 2}, 5 (1964).

\bibitem{suhl}H. Suhl, Phys. Rev. {\bf 138}, A515 (1965).


\bibitem{gruner}G. Gruner and A. Zawadowski, in {\sl Progress in Low Temperature Physics}, edited by D. F. Brewer (North Holland,
Amsterdam, 1978), Vol. {\bf VIIB}, p 591.

\bibitem{martin}R. M. Martin
Phys. Rev. Lett. {\bf 48}, 362-365 (1982).

\bibitem{schraiman}B. I. Shraiman and E. D. Siggia
Phys. Rev. Lett. {\bf 62}, 1564-1567 (1989).

\bibitem{chubukov}B. L. Altshuler , A. V. Chubukov, A. Dashevskii
A, A. M. Finkel'stein  and D. K.  Morr, Europhysics Letters {\bf 41}, 401 (1998). 

\bibitem{coleman}P. Coleman,
C. P\' epin,Q. Si and R. Ramazashvili, J. Cond Matt {\bf 13}, R723 (2001).


\bibitem{senthil}T. Senthil, Matthias Vojta, Subir Sachdev,
Physical Review B {\bf 69}, 035111 (2004)
 

\bibitem{pepin05}C. P\' epin, Phys. Rev. Lett. {\bf 94}, 066402 (2005).


%\bibitem{jones}Jones BA, Varma CM. Critical point in the solution of the two magnetic impurity problem. [Journal Paper] Physical Review B-Condensed Matter, vol.40, no.1, 1 July 1989, pp.324-9.

\bibitem{onuki}S. Kawarazaki et. al. Phys. Rev B {\bf 61}, 4167 (2000).

\bibitem{silke}S.Paschen, T.~L{\"u}hmann, S.~Wirth, P.~Gegenwart,
O.~Trovarelli, Ch.~Geibel, F.~Steglich, P.~Coleman and
Q.~Si, Nature,  Nature {\bf 432}, 881 - 885 (16 December 2004).

\bibitem{affleck}M. Oshikawa, M. Yamanaka, and I. Affleck,
Phys. Rev. Lett. {\bf 78}, 1984 (1997).

\bibitem{oshikawa}M. Oshikawa
Phys. Rev. Lett. {\bf 84}, 3370-3373 (2000)

\bibitem{eliashberg} G. M. Eliashberg, Zh. Eksp. Teor. Fiz. 43, 1005 (1962);
[Sov. Phys.-JETP {\bf 16}, 780 (1963)].

\bibitem{blaizot}J.-P. Blaizot, E. Iancu, and A. Rebhan
Phys. Rev. D {\bf 63}, 065003 (2001).

\bibitem{pethick}G. M. Carneiro and C. J. Pethick, Phys. Rev. {\bf
B11}, 1106 (1975).

\bibitem{kotliar} A. Georges, G. Kotliar,  W. Krauth and M. Rozenberg,
Rev. Mod. Phys. 68, 13 (1996)

\bibitem{parcolletdmft}O. Parcollet, G. Biroli, G. Kotliar, Phys. Rev. B 69, 205108 (2004).

\bibitem{kadanoffbaym} G. Baym and L. P. Kadanoff, Phys. Rev. {\bf{124}}, 287 (1961); G. Baym, Phys. Rev. {\bf{127}}, 1391 (1962).

\bibitem{parcollet97a}O. Parcollet and A. Georges, PRL {\bf 79}, 4665-8
(1997).

\bibitem{powell}Stephen Powell, Subir Sachdev and Hans Peter Buchler,
cond-mat/0502299.

\bibitem{rech}J. Rech, P. Coleman, O. Parcollet and G. Zarand to be
published (2005). 

\bibitem{colemanpepin03}P. Coleman and C. P\' epin, Phys. Rev. B
,Physical Review B {\bf 68} 220405 (2003) .


\bibitem{indranil}P. Coleman \& I. Paul. Phys. Rev. B {\bf{70}},1 (2004).


%\bibitem{nozieres}P. Nozi{\`e}res, Journal de Physique C 37, C1-271, 1976 ;
%P. Nozi{\`e}res and A. Blandin, Journal de Physique 41, 193, 1980.


\bibitem{zouanderson}P. W. Anderson and Z. Zou
Phys. Rev. Lett. {\bf 60}, 132-135 (1988).

\bibitem{ambiguity}To avoid the ambiguities of unit cells,
it is important here that the $\vec{Q}$ vector does not have certain
values. For a  helimagnet, for example, we must avoid the
commensurate case where $\vec{Q }= (\pi/N,\pi/N,\pi/N)$, where $N=2$
is the parameter for an SP (2N) generalization.

\bibitem{potthoff}M. Potthoff, cond-mat/0406671 to be published.

\bibitem{rammer}J. Rammer and H. Smith, Rev. Mod. Phys., {\bf 58}, 323(1986).

\bibitem{kamenev}A. Kamenev, cond-mat/0412296 to be published (05).

\bibitem{clogston61}A. M. Clogston and P. W. Anderson,
Bull. Am. Phys. Soc {\bf{ 6}}, 124 (1961).

\bibitem{compensationnote}For a detailed discussion of this point, see
Appendix A in P. Coleman \& I. Paul. Phys. Rev. B {\bf{70}},1 (2004).

\bibitem{miranda} J. C. Xavier, E. Novais and E. Miranda, Phys. Rev. B
{\bf 65}, 214406 (2002).

\bibitem{pivivarov}E. Pivovarov and Q. Si, cond-mat/0304129. 

\end{thebibliography}
\end{document}